\documentclass[a4paper,11pt]{article}
\pdfoutput=1
\usepackage{jheppub}
\usepackage{textcomp}

\title{Production of the (predicted) $K^*(4307)$ in $B$ decays}
\date{\today}
\author{Xiu-Lei Ren,}
\emailAdd{xiulei.ren@rub.de}
\affiliation{Ruhr-Universit\"{a}t Bochum, Fakult\"{a}t f\"{u}r Physik und Astronomie, Institut f\"{u}r Theoretische Physik II, D-44780 Bochum, Germany.}

\author{K. P. Khemchandani,}
\emailAdd{kanchan.khemchandani@unifesp.br}
\affiliation{Universidade Federal de Sao Paulo, C.P. 01302-907, Sao Paulo, Brazil.}

\author{A. Mart\'inez Torres}
\emailAdd{amartine@if.usp.br}
\affiliation{Universidade de Sao Paulo, Instituto de Fisica, C.P. 05389-970, Sao 
Paulo, Brazil.}

\abstract{In this work we study the production of $K^*(4307)$ in $B$ decays by determining the $J/\psi\pi^{+(0)}K^{0}$ and $J/\psi \pi^- K^+$ invariant mass distributions of the processes $B^+\to J/\psi \pi^+\pi^0 K^0$ and $B^+\to J/\psi \pi^+\pi^- K^+$, respectively. Such $K^*(4307)$ has been recently predicted as a three-body state originated from the dynamics involved in the $K D\bar D^*$ system, with the $KD$ subsystem forming the $D^{*}_{s0}(2317)$ in isospin 0, and the $D\bar D^*$ subsystem generating the $X(3872)$ in isospin 0 and the $Z_c(3900)$ in isospin 1. The hidden charm content of $K^*(4307)$ favors its decay to a state like $J/\psi\pi K$ and the study of $B$-decays with these particles in their final states can constitute a way of finding experimental evidences for such an exotic vector meson, whose width, in spite of its large mass, is still quite narrow (around 18 MeV).}

\begin{document}

\maketitle
\section{Introduction}
In the last years, the $B$-factories have become an unexpected and crucial source of experimental data useful for the understanding of the properties, as well as the discovery, of mesons whose nature seems to challenge the traditional quark model; specially those mesons/baryons with hidden or explicit charm quantum numbers. For instance, the states $D^*_{s0}(2317)$ and $D^*_{s1}(2460)$, observed for the first time in $e^+ e^-$ collisions~\cite{Aubert:2003fg,Besson:2003cp}, were also found in the study of $B\to \bar D D_s\pi$ and $B\to \bar D D^*_s\pi$~\cite{Krokovny:2003zq}, with the particles $D_s\pi$ ($D^*_s\pi$) being originated from the decay $D^*_{s0}(2317)\to D_s\pi$ [$D^*_{s1}(2460)\to D^*_s\pi$]. Such studies were crucial for confirming the quantum numbers of $D^*_{s0}(2317)$ and ruling out the possible spin 0 assignment for $D^*_{s1}(2460)$. The quantum numbers $J^{PC}=1^{++}$ of $X(3872)$ were indeed confirmed by the LHCb collaboration in the study of the decay $B^+\to J/\psi \pi^+\pi^- K^+$, followed by $J/\psi\to \mu^+\mu^-$, where  $X(3872)$ was found in the $\pi^+\pi^- J/\psi$ invariant mass distribution~\cite{Aaij:2013zoa}. The same decay process, as well as the decays $B^+\to J/\psi\pi^+\pi^0 K^0 $ and $B^-\to  J/\psi \pi^-\pi^0K^0$, were previously investigated by the Belle~\cite{Choi:2003ue,Adachi:2008te,Choi:2011fc} and BaBar collaborations~\cite{Aubert:2004ns,Aubert:2004zr} in the context of searching for $X(3872)$ and a possible charged partner. Along the same line, the study of the decay $B\to K\pi^\pm \psi^\prime$ lead to the claim by the Belle collaboration of the existence of a state with a minimal tetraquark configuration, $Z^\pm(4430)$, in the $\pi^\pm \psi^\prime$ invariant mass distribution~\cite{Choi:2007wga,Chilikin:2013tch}. Such state has also been claimed by the LHCb collaboration, which arrived to the conclusion that a highly significant $Z^-(4430)\to \psi^\prime\pi^-$ is needed to describe the decay $B^0\to \psi^\prime \pi^- K^+$~\cite{Aaij:2014jqa}. The experimental observation of baryons with a minimal content of five quarks has also come from the decay of a baryon with bottom quantum number. Particularly, charmonium pentaquark states were claimed by the LHCb collaboration in the $J/\psi p$ invariant mass of the decay process $\Lambda^0_b \to J/\psi K^- p$~\cite{Aaij:2015tga,Aaij:2019vzc}. 

Interestingly, all the above mentioned states share a property: the meson states can be interpreted as tetraquarks or as states obtained from the dynamics involved in two-meson systems, while the baryon states can be understood as pentaquarks or as states originated from meson-baryon systems (for some recent reviews on these topics see, for example, Refs.~\cite{Chen:2016qju,Lebed:2016hpi,Esposito:2016noz,Hosaka:2016pey,Guo:2017jvc,Olsen:2017bmm}). With the amount of data collected from $B$ decays during the past years, it is natural to ask whether there could be signals for other kind of exotic states, like those formed by the interaction of three hadrons, thus, a minimal configuration of six quarks in case of mesons and of seven in case of baryons. In the recent years, formation of three-body bound states/resonances with hidden or explicit charmed has been claimed~\cite{SanchezSanchez:2017xtl,Ma:2017ery,MartinezTorres:2018zbl,Valderrama:2018sap,Ren:2018pcd,Di:2019jsx,Ren:2019umd,Wu:2019vsy,Huang:2019qmw}, however, an experimental investigation of these states seems still not being in the agenda of the facilities around the world. Particularly interesting is the exotic $K^*$ vector meson found in Refs.~\cite{Ma:2017ery,Ren:2018pcd}, a state with hidden charm, a mass around 4300 MeV, but still narrow, with a width of around 18 MeV~\cite{Ren:2018pcd}. As shown in Ref.~\cite{Ren:2018pcd}, such state arises from the dynamics involved in the $K D\bar D^*$ system when the interaction of the $D\bar D^*$ subsystem  generates the $X(3872)$ in isospin 0 and the $Z_c(3900)$ in isospin 1. The $K^*(4307)$, with a dominant $KZ_c(3900)$ component in its wave function, can naturally decay to a final state formed by $K J/\psi\pi$, with the $J/\psi$ and $\pi$ coming from the decay of $Z_c(3900)$. In this way, an experimental reconstruction of the $J/\psi\pi K$ invariant mass could confirm the existence of such an excited $K^*$ state, where its narrow width should help in its identification. And the fact that information on this invariant mass could be obtained from the existing experimental data on $B^\pm\to J/\psi\pi^\pm\pi^0 K^0$ or $B^+\to J/\psi \pi^+\pi^- K^+$ is especially motivating.

Conducting such experimental research could even open a whole new era on the hunting for exotic states, since the last excited state of a $K$/ $K^*$ observed experimentally according to the Particle Data Group is a Kaon whose mass is around 3100 MeV~\cite{Tanabashi:2018oca}. There is then a vast energy region in which the formation of exotic $K/K^*$ states has been totally unexplored. Having this in mind, in this work, we determine the branching ratio for the processes $B^+\to J/\psi \pi^0 \pi^+ K^0$, through $B^+\to \pi^{0(+)} K^{*+(0)}(4307)\to \pi^{0(+)} K^0 Z^{+(0)}_c(3900)\to \pi^{0(+)} K^0J/\psi \pi^{+(0)}$, and $B^+\to J/\psi\pi^+\pi^- K^+$, through $B^+\to \pi^+ K^{*0}(4307)\to \pi^+ K^+Z^-_c(3900)\to \pi^+ K^+J/\psi\pi^- $, and reconstruct the $J/\psi \pi^{+(0)} K^0$ and $J/\psi\pi^- K^+$ invariant mass distributions with the purpose of studying the $K^*(4307)$ signal in them.

\section{Formalism}
\begin{figure}[h!]
\includegraphics[width=\textwidth]{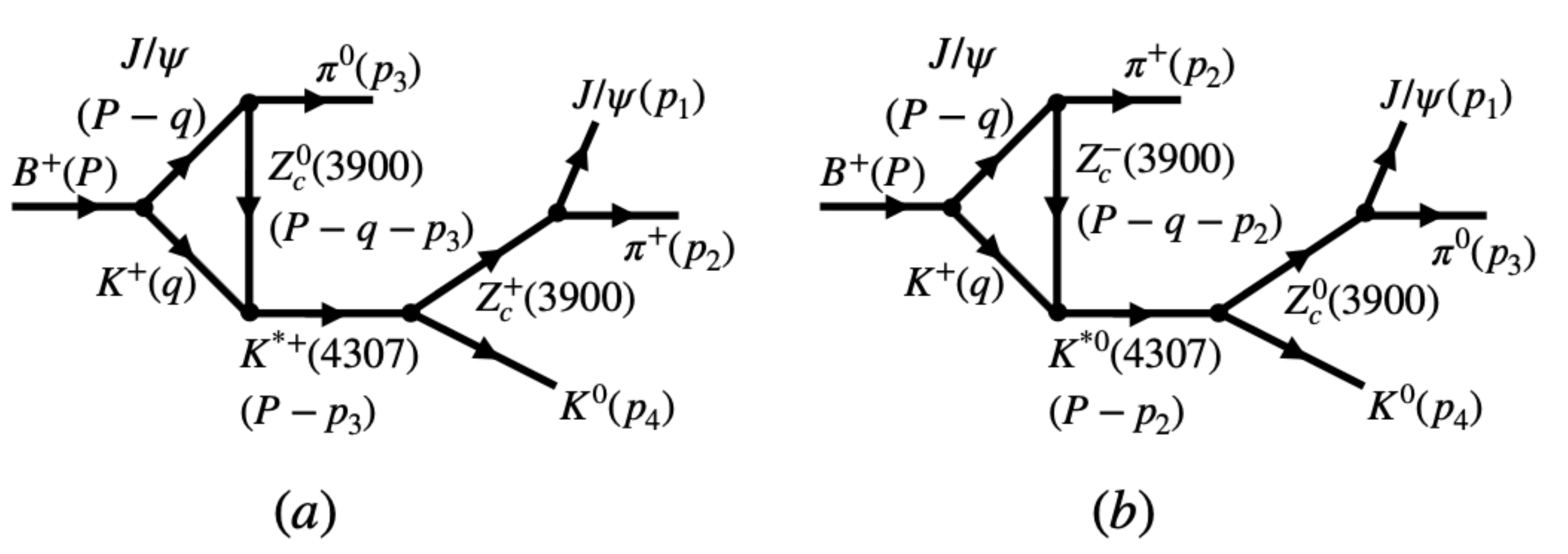}
\caption{Diagrammatical representation of the decay process $B^+\to J/\psi\pi^+\pi^0 K^0$ through formation of $K^{*+\,(0)}(4307)$.}\label{triangular}
\end{figure}

\begin{figure}[h!]
\centering
\includegraphics[width=0.5\textwidth]{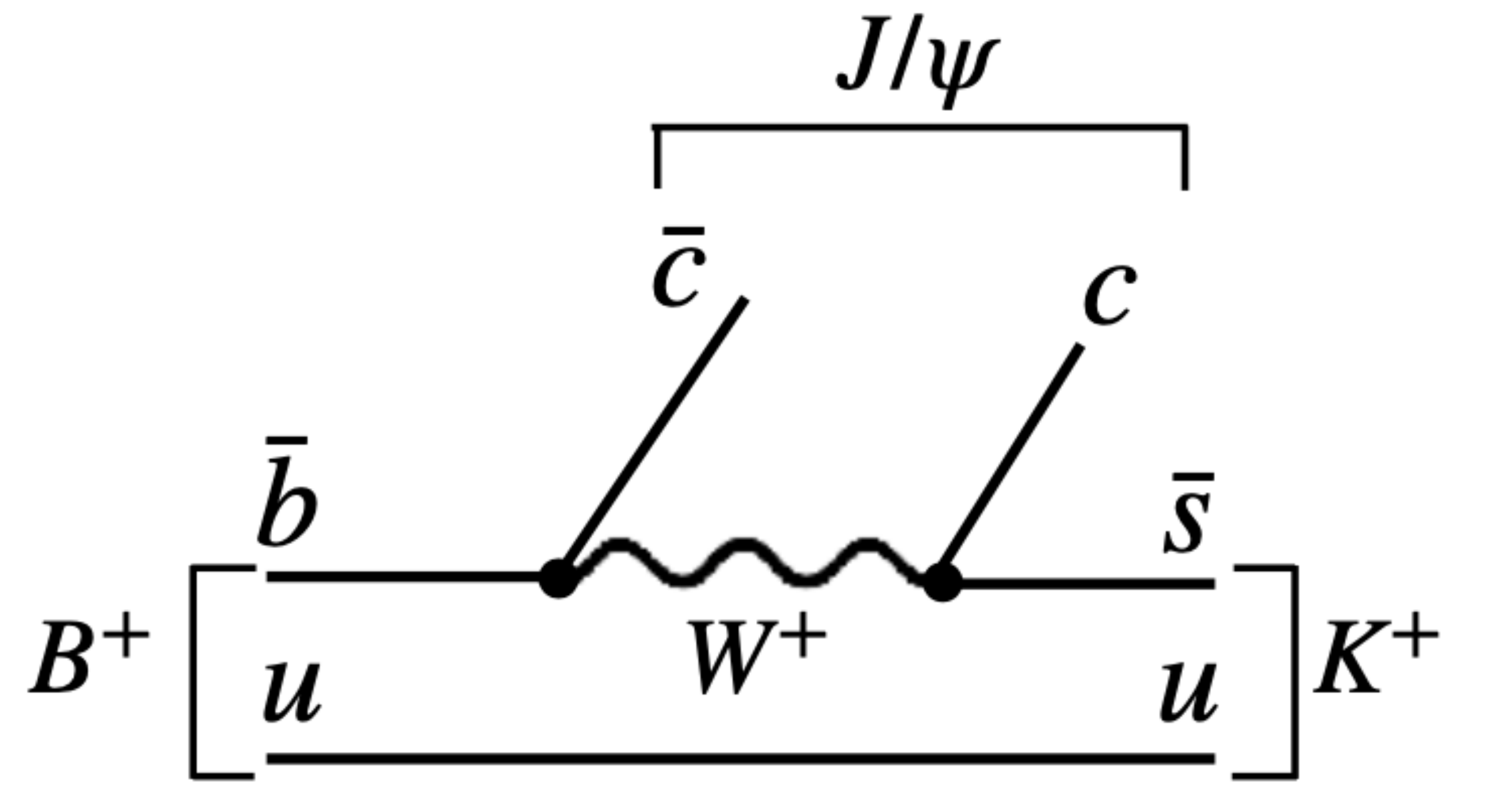}
\caption{Weak vertices involved in the decay of a $B^+$ into a $J/\psi$ and a $K^+$.}\label{Bquarks}
\end{figure}

The decay process $B^+\to J/\psi\pi^+\pi^0 K^0$ proceeding through $K^*(4307)$ formation can be visualized diagrammatically as shown in Fig.~\ref{triangular}, where the interaction between a $K^+$ and a $Z^{0\,(-)}_c(3900)$ generates the $K^{*+\,(0)}(4307)$~\cite{Ren:2018pcd}, which decays to $J/\psi\pi^{+\,(0)} K^0$. The nature of $Z_c(3900)$ is still under debate. Here, as done in Ref.~\cite{Ren:2018pcd}, we follow the model of Ref.~\cite{Aceti:2014uea} where the state is generated from the interaction between $D\bar D^*$ and $J/\psi \pi$ within coupled channels as a weakly bound state of the $D\bar D^*$ system, with a finite width from its decay to the $J/\psi\pi$ channel.
Due to the nature of $K^*(4307)$ and $Z_c(3900)$, the weak vertex $B^+\to J/\psi K^+$  is the most favored for forming $Z_c(3900)$ and $K^*(4307)$. At the quark level, it involves internal emission of a $W^+$ via $\bar b\to \bar c$ ($W^+\to c\bar s)$ transitions, which are both Cabibbo favored (see Fig.~\ref{Bquarks}).
Based on the quantum chromodynamics factorization approach for non-leptonic $B$-meson decays~\cite{Beneke:2000ry}, the amplitude related to the weak vertex shown in Fig.~\ref{Bquarks} can be written as
\begin{align}
t_{B^+\to J/\psi K^+}=\frac{G_F}{\sqrt{2}}V_{cb}V^*_{cs}a_2\langle J/\psi|(\bar c c)_V|0\rangle\langle K|(\bar b s)_V| B\rangle,\label{tB}
\end{align}
where $G_F$ is the Fermi coupling constant, $V_{cb}$, $V^*_{cs}$ are elements of the Cabibbo-Kobayashi-Maskawa matrix, $a_2$ is an effective coupling constant, $\langle J/\psi|(\bar c c)_V|0\rangle$ is the factorized amplitude for the production of a $J/\psi$ via the vector current $\bar c\gamma_\mu c$ and $\langle K|(\bar b s)_V| B\rangle$ represents the transition matrix element $B^+\to K^+$. The amplitude $\langle J/\psi|(\bar c c)_V|0\rangle$ can be parametrized in terms of the decay constant $f_{J/\psi}$, the mass $m_{J/\psi}$ and the polarization vector $\epsilon_{J/\psi\,\mu}$ of the $J/\psi$ vector meson as~\cite{Neubert:1997uc}
\begin{align}
\langle J/\psi|(\bar c c)_V|0\rangle=\epsilon_{J/\psi\,\mu} m_{J/\psi}f_{J/\psi},
\end{align}
while the transition matrix element $B^+(p)\to K^+(p^\prime)$ can be written as~\cite{Neubert:1997uc}
\begin{align}
\langle K^+(p^\prime)|(\bar b s)_V| B^+(p)\rangle=\left[(p+p^\prime)_\mu-\frac{m^2_{B^+}-m^2_{K^+}}{Q^2}Q_\mu\right]F_1(Q^2)+\frac{m^2_{B^+}-m^2_{K^+}}{Q^2}Q_\mu F_0(Q^2).\label{tran}
\end{align}
In Eq.~(\ref{tran}), $Q_\mu=(p-p^\prime)_\mu$, $F_1(Q^2)$ and $F_0(Q^2)$ correspond to form factors, which satisfy the condition $F_1(0)=F_0(0)$~\cite{Neubert:1997uc}, and $m_{B^+}$ ($m_{K^+}$) is the mass of the $B^+$ ($K^+$) meson. The $Q^2$ dependence of these form factors can be written as~\cite{Deandrea:1993ma}
\begin{align}
F_i(Q^2)=\frac{F_i(0)}{1-Q^2/m^2_{P\,i}},
\end{align}
with $i=0,1$, $m_{P\,i}$ being the mass of the lowest lying meson with the appropriate quantum numbers, i.e., $J^P=0^+$ for $F_0$ ($m_P=5890$ MeV) and $1^-$ for $F_1$ ($m_P=5430$ MeV), and
\begin{align}
F_1(0)=F_0(0)=0.49\pm 0.12.
\end{align}
Using Eq.~(\ref{tB}), we can determine the branching ratio for the process $B^+\to J/\psi K^+$ as
\begin{align}
\text{Br}(B^+\to J/\psi K^+)=\frac{|\vec{p}_\text{CM}|}{8\pi \Gamma_{B^+}m^2_{B^+}}\sum_\lambda |t_{B^+\to J/\psi K^+}|^2,
\end{align}
where the symbol $\sum\limits_\lambda$ indicates sum over the polarizations of $J/\psi$,  
$|\vec{p}_\text{CM}|$ is the center of mass momentum of the $J/\psi K^+$ system, $\Gamma_{B^+}$ is the width of the $B^+$ meson 
and, from Eqs.~(\ref{tB}) and (\ref{tran}) 
\begin{align}
\sum_\lambda |t_{B^+\to J/\psi K^+}|^2&=\frac{G^2_F}{2}|V_{cb}|^2|V_{cs}|^2|a_2|^2 m^2_{J/\psi}f^2_{J/\psi} F^2_1(Q^2)\nonumber\\
&\quad\times\left[-(p+p^\prime)^2+\frac{(m^2_{B^+}-m^2_{K^+})^2}{m^2_{J/\psi}}\right].\label{sumL}
\end{align}
Considering $a_2=0.21\pm 0.02$~\cite{Neubert:1997uc}, $G_F=1.166\times 10^{-11}$ MeV${}^{-2}$, $|V_{cs}|=0.977\pm 0.017$, $|V_{cb}|=(42.2\pm 0.8)\times 10^{-3}$, $m_{J/\psi}=3096.9\pm 0.006$ MeV, $\Gamma_{B^+}=(4.01839\pm0.0098)\times 10^{-10}$ MeV~\cite{Tanabashi:2018oca}, and $f_{J/\psi}=405\pm 14$ MeV~\cite{Neubert:1997uc}, we get 
\begin{align}
\text{Br}(B^+\to J/\psi K^+)\simeq (0.83\pm0.32)\times 10^{-3},
\end{align}
which is compatible with the measured branching ratio of~\cite{Tanabashi:2018oca}
\begin{align}
\text{Br}(B^+\to J/\psi K^+)_\text{measured}=(1.01\pm 0.028)\times 10^{-3}.\label{Bexp} 
\end{align}

Based on the above discussion, the dominant contribution from the weak vertex in the processes depicted in Fig.~\ref{triangular}, can be written, for convenience, as
\begin{align}
t_{B^+\to J/\psi K^+}=C_{B^+\to J/\psi K^+}(P+q)\cdot\epsilon_{J/\psi}(P-q),
\end{align}
where $P^\mu$ ($q^\mu$) is the four-momentum of the $B^+$ ($K^+$), and the coefficient $C_{B^+\to J/\psi K^+}$, which corresponds to $\frac{G_F}{\sqrt{2}}|V_{cb}||V_{cs}||a_2| m_{J/\psi}f_{J/\psi} F_1$ of Eq.~(\ref{sumL}), is fixed to reproduce the observed branching ratio, i.e., Eq.~(\ref{Bexp}), to be more in agreement with the experimental finding,
\begin{align} 
C_{B^+\to J/\psi K^+}=(7.16\pm0.11)\times 10^{-8}.\label{CB}
\end{align}

Since $Z_c(3900)$ and $K^*(4307)$ couple to $J/\psi\pi$ and $K Z_c(3900)$, respectively, in s-wave~\cite{Aceti:2014uea,Ren:2018pcd}, we can introduce the coupling constants  $g_{Z_c\to J/\psi\pi}$ and $g_{K^*\to K Z_c}$ to describe the contribution from these vertices in Fig.~\ref{triangular}. Such contribution can be expressed in terms of the contraction $\epsilon\cdot \epsilon^\prime$ between the polarization vectors of the particles involved, $Z_c(3900)$ and $J/\psi$ or $K^*(4307)$ and $Z_c(3900)$, and the corresponding coupling constant, obtained from Refs.~\cite{Aceti:2014uea,Ren:2018pcd}. In this way, using the Feynman rules, the amplitudes associated with the diagrams in Fig.~\ref{triangular} are given by
\begin{align}
-i t_\mathcal{A}&=C_\mathcal{A} \frac{1}{s_{3(\mathcal{A})}-m^2_{K^*(\mathcal{A})}+i \Gamma_{K^*(\mathcal{A})}m_{K^*(\mathcal{A})}}\frac{1}{s_{2(\mathcal{A})}-m^2_{Z_{\mathcal{A}}}+i\Gamma_{Z(\mathcal{A})}m_{Z(\mathcal{A})}}\nonumber\\
&\quad\times\Bigg[F_-\Bigg\{P_\sigma-\frac{(P-p_\mathcal{A})_\sigma}{M^2_{Z(\mathcal{A})}}(P^2-p_\mathcal{A}\cdot P)\Bigg\}I^{(0\mathcal{A})}+\left\{F_++\frac{F_-}{M^2_{Z(\mathcal{A})}}\left(P^2-p_\mathcal{A}\cdot P\right)\right\}I^{(1\mathcal{A})}_\sigma\nonumber\\
&\quad\quad\quad+\frac{(P-p_\mathcal{A})_\sigma}{M^2_{Z(\mathcal{A})}}\times\left\{F_+p^\mu_\mathcal{A}-2\frac{P^2}{m^2_{J/\psi}}P^\mu\right\}I^{(1\mathcal{A})}_\mu+\Bigg\{\frac{P_\sigma}{m^2_{J/\psi}}+\frac{(P-p_\mathcal{A})_\sigma}{M^2_{Z(\mathcal{A})}}\nonumber\\
&\quad\quad\quad\times\left(1+\frac{p_\mathcal{A}\cdot P}{m^2_{J/\psi}}\right)\Bigg\}I^{(2\mathcal{A})}-\frac{1}{M^2_{Z(\mathcal{A})}}\Bigg\{F_+p^\mu_\mathcal{A}-2\frac{P^2}{m^2_{J/\psi}}P^\mu\Bigg\}I^{(2\mathcal{A})}_{\sigma\mu}\nonumber\\
&\quad\quad\quad+\frac{(P-p_\mathcal{A})_\sigma}{M^2_{Z(\mathcal{A})}}\frac{(-p^\mu_\mathcal{A}+2 P^\mu)}{m^2_{J/\psi}}I^{(3\mathcal{A})}_\mu-\Bigg\{\frac{1}{m^2_{J/\psi}}+\frac{1}{M^2_{Z(\mathcal{A})}}\left(1+\frac{p_\mathcal{A}\cdot P}{m^2_{J/\psi}}\right)\Bigg\}I^{(3\mathcal{A})}_\sigma\nonumber\\
&\quad\quad\quad-\frac{1}{M^2_{Z(\mathcal{A})} m^2_{J/\psi}}\Bigg\{(P-p_\mathcal{A})_\sigma I^{(4\mathcal{A})}+(-p^\mu_\mathcal{A}+2 P^\mu)I^{(4\mathcal{A})}_{\sigma\mu}-I^{(5\mathcal{A})}_\sigma\Bigg\}\Bigg]\nonumber\\
&\quad\times\left[-g^{\sigma\rho}+\frac{(P-p_\mathcal{A})^\sigma(P-p_\mathcal{A})^\rho}{m^2_{K^*(\mathcal{A})}}\right]\left[-g_{\rho\beta}+\frac{(p_1+p_2)_\rho p_{2\beta}}{m^2_{Z(\mathcal{A})}}\right]\epsilon^\beta_{J/\psi}(p_1),\label{tA}
\end{align}
where the subscript $\mathcal{A}$ refers to the diagrams $(a)$ and $(b)$ in Fig.~\ref{triangular}, with $s_{3(\mathcal{A})}=s_{124}=(p_1+p_2+p_4)^2$~[$s_{134}=(p_1+p_3+p_4)^2$], $s_{2(\mathcal{A})}=s_{12}=(p_1+p_2)^2$~[$s_{13}=(p_1+p_3)^2$], $p_\mathcal{A}=p_3$~($p_2$), $m_{Z(\mathcal{A})}=m_{Z^+_c}$~($m_{Z^0_c}$), $M_{Z(\mathcal{A})}=m_{Z^0_c}$~($m_{Z^-_c}$) is the mass of the $Z_c$ particle involved in the triangular loops, $m_{K^*(\mathcal{A})}=m_{K^{*+}(4307)}$~[$m_{K^{*0}(4307)}$], $\Gamma_{Z(\mathcal{A})}=\Gamma_{Z^+_c}$~($\Gamma_{Z^0_c}$), $\Gamma_{K^*(\mathcal{A})}=\Gamma_{K^{*+}(4307)}$~[$\Gamma_{K^{*0}(4307)}$] for the diagram $(a)$ [$(b)$]. In Eq.~(\ref{tA}), 
\begin{align}
F_\pm=1\pm\frac{P^2}{m^2_{J/\psi}},
\end{align}
and the constant $C_\mathcal{A}$ corresponds to a product of the coupling constants involved in the different vertices shown in Fig.~\ref{triangular}. To be more exact, 
\begin{align}
C_{a}=-C_{b}=\frac{\sqrt{2}}{3}C_{B^+\to J/\psi K^+}g^2_{Z_c\to J/\psi\pi}g^2_{K^*\to KZ_c},
\end{align}
where $g_{Z_c\to J/\psi\pi}$ is the coupling constant of $Z_c(3900)$ to the $J/\psi\pi$ system in isospin 1, and whose value is obtained from the model of Ref.~\cite{Aceti:2014uea}, and $g_{K^*\to KZ_c}$ represents the coupling of $K^*(4307)$ to a $KZ_c(3900)$ system in isospin $1/2$, which can be found from the model of Ref.~\cite{Ren:2018pcd}. The values used here are $g_{Z_c\to J/\psi\pi}=3715$ MeV and $g_{K^*\to KZ_c}=22143$ MeV~\cite{Ren:2019umd}. The phase convention $|\pi^+>=-|I=1,I_3=1\rangle$, $|K^-\rangle=-|I=1/2,I_3=-1/2\rangle$ has been used in our calculations.

In Eq.~(\ref{tA}), $I^{(0\mathcal{A})}$, $I^{(1\mathcal{A})}_\alpha$, $I^{(2\mathcal{A})}_{\alpha\beta}$, $I^{(3\mathcal{A})}_\alpha$, $I^{(4\mathcal{A})}_{\alpha\beta}$ and $I^{(5\mathcal{A})}_\alpha$ are integrals defined as 
\begin{align}
I^{(0\mathcal{A})}&=\int\frac{d^4 q}{(2\pi)^4}\frac{1}{\mathcal{D}(q,p_\mathcal{A})},\quad I^{(1\mathcal{A})}_\alpha=\int\frac{d^4 q}{(2\pi)^4}\frac{q_\alpha}{\mathcal{D}(q,p_\mathcal{A})},\quad I^{(2\mathcal{A})}_{\alpha\beta}=\int\frac{d^4 q}{(2\pi)^4}\frac{q_\alpha q_\beta}{\mathcal{D}(q,p_\mathcal{A})},\nonumber\\
I^{(3\mathcal{A})}_\alpha&=\int\frac{d^4 q}{(2\pi)^4}\frac{q^2q_\alpha}{\mathcal{D}(q,p_\mathcal{A})},\quad I^{(4\mathcal{A})}_{\alpha\beta}=\int\frac{d^4 q}{(2\pi)^4}\frac{q^2q_\alpha q_\beta}{\mathcal{D}(q,p_\mathcal{A})},\quad I^{(5\mathcal{A})}_\alpha=\int\frac{d^4 q}{(2\pi)^4}\frac{q^4 q_\alpha}{\mathcal{D}(q,p_\mathcal{A})},\label{int}
\end{align}
where
\begin{align}
\mathcal{D}(q,p_\mathcal{A})=[(P-q)^2-m^2_{J/\psi}+i\epsilon][q^2-m^2_{K^+}+i\epsilon][(P-p_\mathcal{A}-q)^2-M^{2}_{Z(\mathcal{A})}+i\epsilon].\label{den}
\end{align}
The integrals $I^{(2\mathcal{A})}$ and $I^{(4\mathcal{A})}$ in Eq.~(\ref{tA}) correspond to the contraction of the metric tensor $g^{\alpha\beta}$ with the integrals $I^{(2\mathcal{A})}_{\alpha\beta}$ and $I^{(4\mathcal{A})}_{\alpha\beta}$ of Eq.~(\ref{int}), respectively. The integrals in Eq.~(\ref{int}) are regularized by using a cut-off of $\sim 700$ MeV for the center of mass momentum of the $K-Z_c$ system, which is compatible with the cut-off used in Ref.~\cite{Ren:2018pcd} to generate the $K^*(4307)$ from the $KD\bar D^*$ system. For more details on the calculation of the integrals in Eq.~(\ref{int}) we refer the reader to the Appendix~\ref{ap}. 

Using the amplitude in Eq.~(\ref{tA}), the decay width $\Gamma$ for the process $B^+ (P)\to J/\psi(p_1) \pi^+(p_2) \pi^0(p_3) K^0(p_4)$ can be obtained as
\begin{align}
\Gamma&=\frac{1}{2 m_{B^+}}\int\frac{d^3p_1}{(2\pi)^3 2E_1(\vec{p}_1)}\int\frac{d^3p_2}{(2\pi)^3 2E_2(\vec{p}_2)}\int\frac{d^3p_3}{(2\pi)^3 2E_3(\vec{p}_3)}\int\frac{d^3p_4}
{(2\pi)^3 2E_4(\vec{p}_4)}\nonumber\\
&\quad\quad\quad\quad\quad\times (2\pi)^4\delta^{(4)}(P-p_1-p_2-p_3-p_4)\sum_\lambda|t_a+t_b|^2.\label{width}
\end{align}
Using the $\delta$-function of Eq.~(\ref{width}), and the relations
\begin{align}
s_{124}&=(p_1+p_2+p_4)^2=(P-p_3)^2=m^2_{B^+}+m^2_3-2 m_{B^+} E_3, \nonumber\\
s_{134}&=(p_1+p_3+p_4)^2=(P-p_2)^2=m^2_{B^+}+m^2_2-2m_{B^+}E_2,\label{ss}
\end{align}
with $E_2=\sqrt{\vec{p}^{\,2}_2+m^2_2}$ ($E_3=\sqrt{\vec{p}^{\,2}_3+m^2_3}$) being the energy related to the particle with three-momentum $\vec{p}_2$ ($\vec{p}_3$) and mass $m_2$ ($m_3$) in the rest frame of the decaying particle, we can write Eq.~(\ref{width}) as
\begin{align}
\Gamma&=\frac{1}{(2\pi)^7 2^6 m^3_{B^+}}\int\limits_{s^\text{min}_{124}}^{s^\text{max}_{124}}ds_{124}\int\limits_{s^\text{min}_{134}}^{s^\text{max}_{134}}ds_{134}\int\limits_{E^\text{min}_1}^{E^\text{max}_1} dE_1\int\limits_0^{2\pi} d\phi_1\int\limits_{-1}^{1} d\text{cos}\theta_3\int\limits_{0}^{2\pi}d\phi_3\nonumber\\
&\quad\times\frac{|\vec{p}_2||\vec{p}_3|}{|\vec{p}_2+\vec{p}_3|}\Theta(1-\text{cos}^2\theta_1)\Theta(m_{B^+}-E_1-E_2-E_3)\sum_\lambda|t_a+t_b|^2.\label{Gamma2}
\end{align}
To arrive to Eq.~(\ref{Gamma2}) we have considered, without lost of generality, that particle 2 is along the $z$-axis, and we integrate on the solid angle of particles 1 and 3. The value $\text{cos}\theta_1$, with $\theta_1$ being the angle between the vectors $\vec{p}_1$ and $\vec{p}_2+\vec{p}_3$, is fixed by the $\delta$-function of Eq.~(\ref{width}), 
\begin{align}
\text{cos}\theta_1=\frac{(m_{B^+}-E_1-E_2-E_3)^2-\vec{p}^{\,2}_1-m^2_4-(\vec{p}_2+\vec{p}_3)^2}{2|\vec{p}_1||\vec{p}_2+\vec{p}_3|}.
\end{align}
The related Heaviside $\Theta$-function in Eq.~(\ref{Gamma2}) guaranties that $|\text{cos}\theta_1|\leqslant 1$, as it should be. In this way,  if $(\theta_{23},\phi_{23})$ are the polar and azimutal angles, respectively, that the vector $\vec{p}_2+\vec{p}_3$ forms with the axes we can obtained $\vec{p}_1$ as
\begin{align}
\vec{p}_1=R_z(\phi_{23})R_y(\theta_{23})|\vec{p}_1|\left(\begin{array}{c}\text{sin}\theta_1\text{cos}\phi_1\\\text{sin}\theta_1\text{sin}\phi_1\\\text{cos}\theta_1\end{array}\right),
\end{align}
where $\text{sin}\theta_1=+\sqrt{1-\text{cos}^2\theta_1}$, $|\vec{p}_1|=\sqrt{E^2_1-m^2_1}$, $R_y$ and $Rz$ represent rotation matrices around the $y$ and $z$ axis, respectively,
\begin{align}
R_y(\theta_{23})=\left(\begin{array}{ccc}\text{cos}\theta_{23}&0&\text{sin}\theta_{23}\\0&1&0\\-\text{sin}\theta_{23}&0&\text{cos}\theta_{23}\end{array}\right),\quad
R_z(\phi_{23})=\left(\begin{array}{ccc}\text{cos}\phi_{23}&-\text{sin}\phi_{23}&0\\\text{sin}\phi_{23}&\text{cos}\phi_{23}&0\\0&0&1\end{array}\right).
\end{align}
The angles $\theta_{23}$ and $\phi_{23}$ can be obtained in terms of the components $x$, $y$ and $z$ of the vectors $\vec{p}_2$ and $\vec{p}_3$, which are given by
\begin{align}
\vec{p}_2=|\vec{p}_2|\left(\begin{array}{c}0\\0\\1\end{array}\right),\quad\vec{p}_3=|\vec{p}_3|\left(\begin{array}{c}\text{sin}\theta_3\text{cos}\phi_3\\\text{sin}\theta_3\text{sin}\phi_3\\\text{cos}\theta_3\end{array}\right),
\end{align}
as
\begin{align}
\text{cos}\theta_{23}=\frac{p_{2z}+p_{3z}}{|\vec{p}_2+\vec{p}_3|},&\quad \text{sin}\theta_{23}=+\sqrt{1-\text{cos}^2\theta_{23}},\\
\text{cos}\phi_{23}=\frac{p_{2x}+p_{3x}}{|\vec{p}_2+\vec{p}_3|\text{sin}\theta_{23}},&\quad \text{sin}\phi_{23}=\frac{p_{2y}+p_{3y}}{|\vec{p}_2+\vec{p}_3|\text{sin}\theta_{23}}.
\end{align}
Note that $|\vec{p}_2|=\sqrt{E^2_2-m^2_2}$ and $|\vec{p}_3|=\sqrt{E^2_3-m^2_3}$ are related to the invariant masses $s_{134}$ and $s_{124}$ through Eq.~(\ref{ss}). The momentum $\vec{p}_4$ can be obtained as $\vec{p}_4=-\vec{p}_1-\vec{p}_2-\vec{p}_3$. In this way, the last ingredient to determine Eq.~(\ref{width}) is the limits of the integrals, which are
\begin{align}
s^\text{min}_{124}&=(m_1+m_2+m_4)^2,\quad s^\text{max}_{124}=(m_{B^+}-m_3)^2,\\
s^\text{min}_{134}&=(m_1+m_3+m_4)^2,\quad s^\text{max}_{134}=(m_{B^+}-m_2)^2,\\
E^\text{min}_1&=m_1,\quad E^\text{max}_1=\frac{m^2_{B^+}+m^2_1-s^\text{min}_{234}}{2m_{B^+}},\quad s^\text{min}_{234}=(m_2+m_3+m_4)^2.
\end{align}
From Eq.~(\ref{Gamma2}), we can determine the corresponding $J/\psi\pi K$ invariant mass distribution as
\begin{align}
\frac{d\Gamma}{ds_{124}}&=\frac{1}{(2\pi)^7 2^6 m^3_{B^+}}\int\limits_{s^\text{min}_{134}}^{s^\text{max}_{134}}ds_{134}\int\limits_{E^\text{min}_1}^{E^\text{max}_1} dE_1\int\limits_0^{2\pi} d\phi_1\int\limits_{-1}^{1} d\text{cos}\theta_3\int\limits_{0}^{2\pi}d\phi_3\nonumber\\
&\quad\times\frac{|\vec{p}_2||\vec{p}_3|}{|\vec{p}_2+\vec{p}_3|}\Theta(1-\text{cos}^2\theta_1)\Theta(m_{B^+}-E_1-E_2-E_3)\sum_\lambda|t_a+t_b|^2,\label{Gamma124}\\
\frac{d\Gamma}{ds_{134}}&=\frac{1}{(2\pi)^7 2^6 m^3_{B^+}}\int\limits_{s^\text{min}_{124}}^{s^\text{max}_{124}}ds_{124}\int\limits_{E^\text{min}_1}^{E^\text{max}_1} dE_1\int\limits_0^{2\pi} d\phi_1\int\limits_{-1}^{1} d\text{cos}\theta_3\int\limits_{0}^{2\pi}d\phi_3\nonumber\\
&\quad\times\frac{|\vec{p}_2||\vec{p}_3|}{|\vec{p}_2+\vec{p}_3|}\Theta(1-\text{cos}^2\theta_1)\Theta(m_{B^+}-E_1-E_2-E_3)\sum_\lambda|t_a+t_b|^2.\label{Gamma134}
\end{align}

\begin{figure}
\centering
\includegraphics[width=0.5\textwidth]{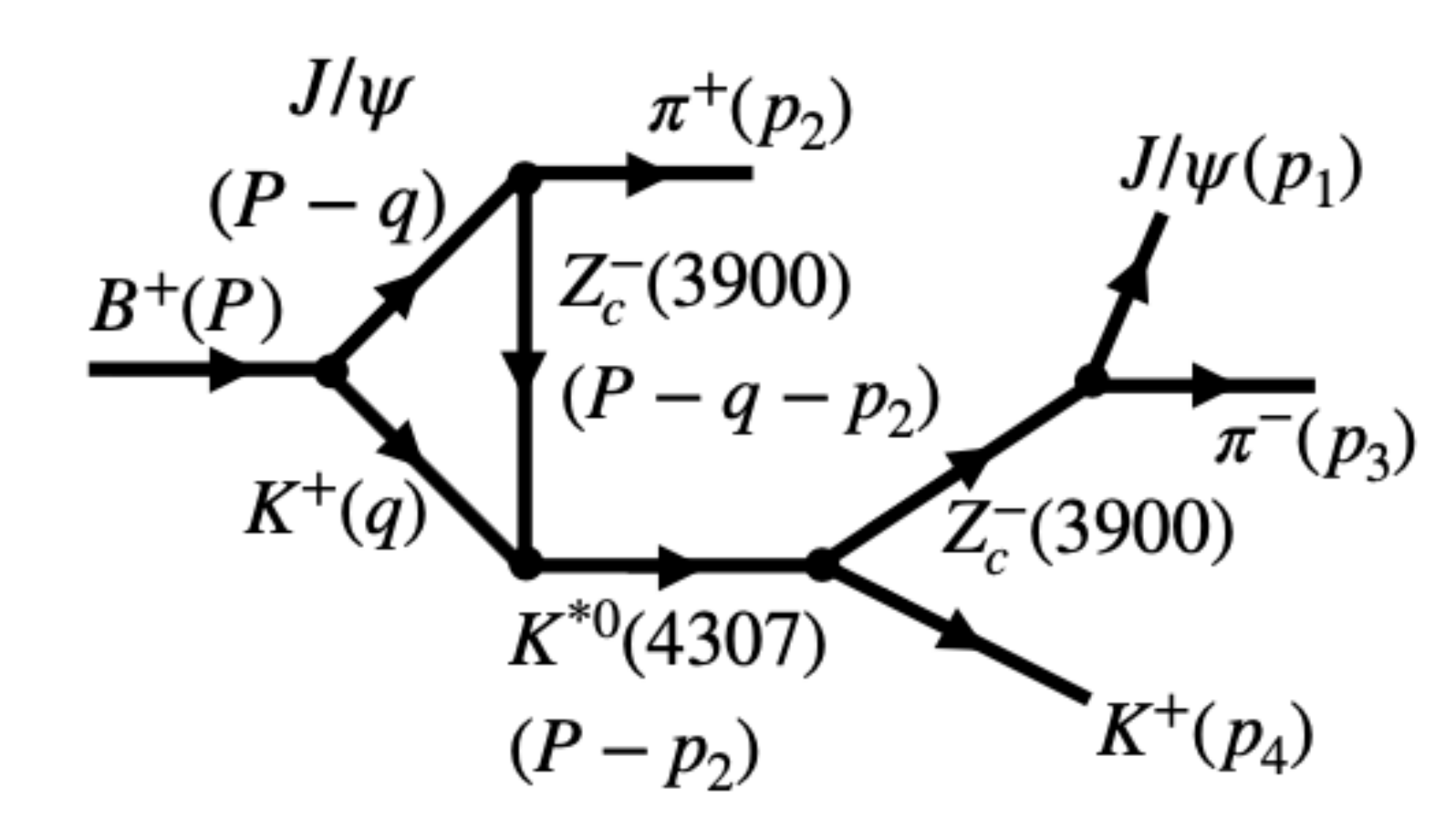}
\caption{Diagrammatical representation of the decay $B^+\to J/\psi \pi^+\pi^- K^+$ through $K^*(4307)$ formation.}\label{triangular2}
\end{figure}

As we mentioned in the introduction, the decay $B^+\to J/\psi \pi^+\pi^-K^+$ has been used for the experimental investigation of the properties of $X(3872)$. In this reaction, the reconstruction of the $J/\psi \pi^- K^+$ invariant mass distribution, can also serve to investigate the properties of $K^*(4307)$. For the process $B^+\to J/\psi \pi^+\pi^-K^+$, as can be seen in Fig.~\ref{triangular2}, the formation of $K^*(4307)$ is completely analogous to the one shown in Fig.~\ref{triangular}(b), with the exception that the vertices $K^{*0}(4307)\to Z^0_c(3900) K^0\to J/\psi \pi^0 K^0$ should be replaced by $K^{*0}(4307)\to Z^-_c(3900) K^+\to J/\psi \pi^- K^+$. This makes that the product of the coupling constants $g_{K^{*0}(4307)\to Z^0_c(3900)}g_{K^0 Z^0_c\to J/\psi \pi^0}$ appearing in the amplitude related to the diagram in Fig.~\ref{triangular}(b) should be substituted by $g_{K^{*0}(4307)\to K^+ Z^-_c(3900)}g_{Z^-_c\to J/\psi \pi^-}$, which, by using the corresponding Clebsch-Gordan coefficients, is $\sqrt{2}$ times bigger than the former product. In this way, the calculation of the decay width for $B^+\to J/\psi \pi^+\pi^- K^+$ and the determination of the $J/\psi \pi^- K^+$ invariant mass distribution is completely analogous to the one for the reaction $B^+\to J/\psi\pi^+\pi^0 K^0$, but we have now contribution from only one Feynman diagram instead of two (see Fig.~\ref{triangular2}) and the couplings, as explained above, are different.

\section{Results}
\begin{figure}[h!]
\centering
\includegraphics[width=0.7\textwidth]{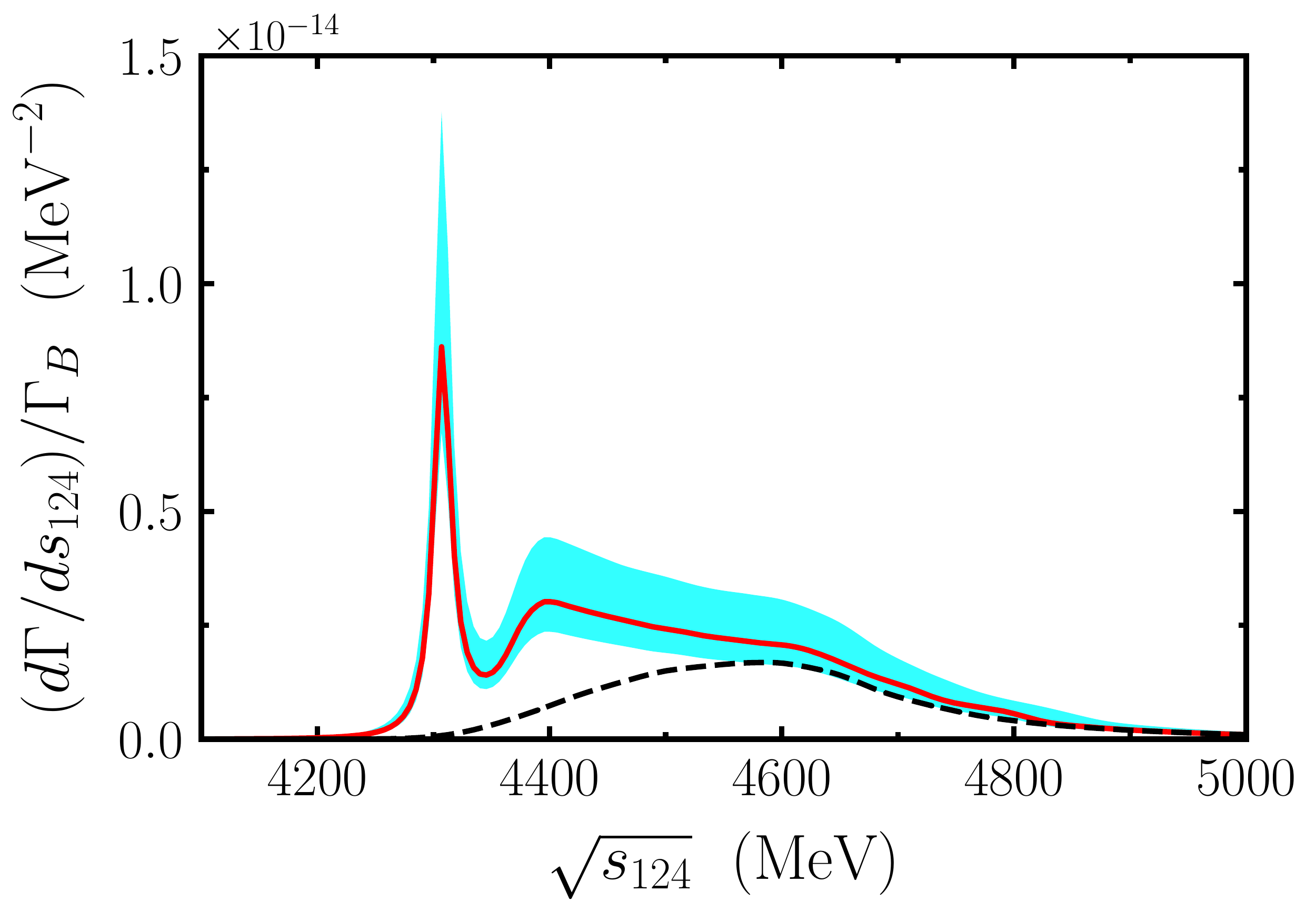}
\caption{Invariant mass distribution, divided by the full width of the $B^+$ meson, as a function of the invariant mass of the $J/\psi \pi^+ K^0$ system, i.e., $\sqrt{s_{124}}$ in Fig.~\ref{triangular}. The solid line corresponds to the result found with a cut-off $\Lambda$ of 700 MeV. The dashed line represents the contribution obtained to $d\Gamma/ds_{124}$ from the diagram in Fig.~\ref{triangular}(b). The band represents the uncertainty associated with $d\Gamma/ds_{124}$ when changing the cut-off in the range $700-750$ MeV, the coupling in Eq.~(\ref{CB}) inside the interval compatible with its error  and considering a 10$\%$ error for the couplings of $Z_c(3900)$ to the $J/\psi\pi$ system and that of $K^*(4307)$ to the $KZ_c(3900)$ system.}\label{inv}
\end{figure}

To obtain the $J/\psi \pi^{+(0)} K^0$ invariant mass distributions of the process $B^+\to J/\psi\pi^+\pi^0 K^0$, we have made use of Eqs.~(\ref{Gamma124}) and (\ref{Gamma134}) considering isospin average masses for those particles belonging to the same isospin multiplet. In such a case, there is no difference between the invariant mass distributions of Eqs.~(\ref{Gamma124}) and (\ref{Gamma134}). In Fig.~\ref{inv} we show $d\Gamma/ds_{124}$ for the process $B^+\to J/\psi\pi^+\pi^0K^0$ as a function of the invariant mass of the $J/\psi\pi^{+} K^0$ system, i.e., $\sqrt{s_{124}}$. 
The solid line in Fig.~\ref{inv} represents the result obtained by using a cut-off $\Lambda$ to regularize the integrals in Eq.~(\ref{int}) of 700 MeV for the center of mass momentum of the $K-Z_c$ system (see Appendix~\ref{ap} for more details). As can be seen, a peak around $4307$ MeV, with a width of 18 MeV, is observed in the distribution due to the formation of $K^*(4307)$, followed by an enhancement around the $K-Z_c(3900)$ threshold, a typical effect when triangular loops are involved in the determination of the amplitudes~\cite{Guo:2015umn,Liu:2015fea,Bayar:2016ftu}, as in our case. We also plot in Fig.~\ref{inv} the contribution to $d\Gamma/ds_{124}$ originated from just the diagram of Fig.~\ref{triangular}(b), which produces a background~\footnote{Note that in the diagram of Fig.~\ref{triangular}(b) the $K^*(4307)$ is formed in the $s_{134}$ invariant mass.} (represented as a dashed-line in Fig.~\ref{inv}). By integrating this distribution, we can get the branching ratio for the process $B^+\to \pi^{0(+)} K^{*+(0)}(4307)\to \pi^{0(+)} K^{0(+)} Z^{+(0)}_c(3900)\to\pi^{0(+)} K^0 J/\psi \pi^{+(0)}$, which is $\mathcal{BR}=1.04\times 10^{-8}$. We can also estimate the uncertainty related to this result. To do this, we vary the cut-off $\Lambda$ in the range $700-750$ MeV, as done in Ref.~\cite{Ren:2018pcd}, the coupling in Eq.~(\ref{CB}) in the range allowed by the related error and we associate a 10$\%$ error to the coupling constants of $Z_c(3900)$ to the $J/\psi\pi$ system and of $K^*(4307)$ to the $KZ_c(3900)$ system. We then generate random numbers inside these intervals and obtain the mean value for the branching ratio and the standard deviation. By doing this, we obtain the band shown in Fig.~\ref{inv} and the estimated branching ratio becomes 
\begin{align}
\mathcal{BR}=(1.05\pm 0.2)\times 10^{-8}.
\end{align}

\begin{figure}[h!]
\centering
\includegraphics[width=0.7\textwidth]{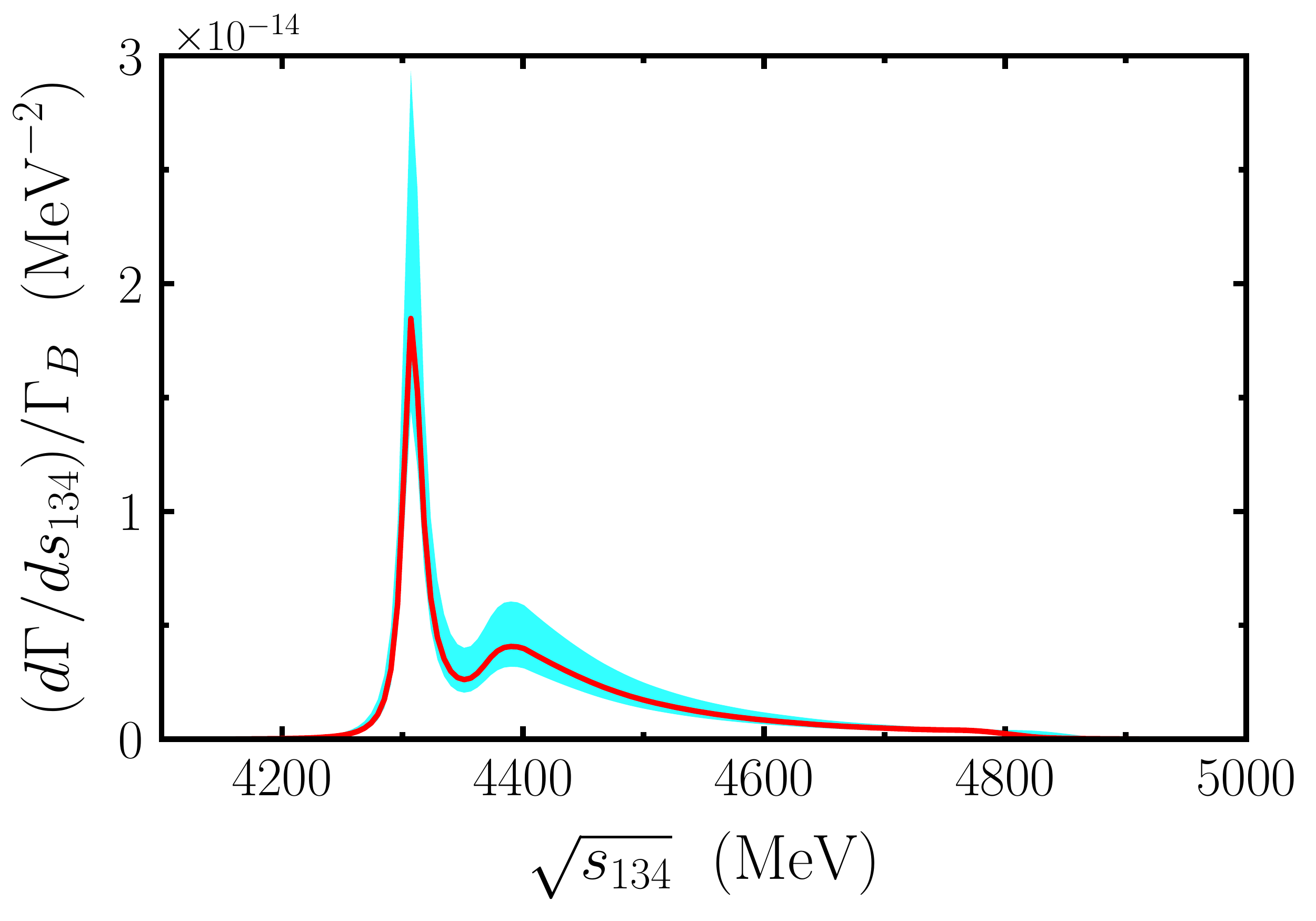}
\caption{Invariant mass distribution, divided by the full width of the $B^+$ meson, as a function of the invariant mass of the $J/\psi \pi^- K^+$ system, i.e., $\sqrt{s_{134}}$ in Fig.~\ref{triangular2}. The solid line and band have the same meaning as that in Fig.~\ref{inv}.}\label{inv2}
\end{figure}

In case of the decay $B^+\to J/\psi \pi^+\pi^- K^+$, the $d\Gamma/ds_{J/\psi \pi^- K^+}$ distribution is shown in Fig.~\ref{inv2} as a function of the $J/\psi \pi^- K^+$ invariant mass, i.e., $\sqrt{s_{134}}$ in Fig.~\ref{triangular2}. As can be seen, a peak structure related to the formation of $K^*(4307)$ is observed, together with an enhancement around 4400 MeV, as in case of Fig.~\ref{inv} and which is related to the threshold of the $K-Z_c$ system. The error band shown in the figure has been obtained in the same way as that of Fig.~\ref{inv} and the solid line represents the result found with a cut-off of 700 MeV.
\section{Conclusion}
By using isospin average masses between the members of the same multiplet, we have determined the $J/\psi \pi^{\pm,0} K^{+,0}$ invariant mass distributions of $B^+\to J/\psi \pi^+\pi^0 K^0$ and $B^+\to J/\psi \pi^+\pi^- K^+$ with the purpose of analyzing the signal related to the formation of $K^{*}(4307)$.  We find that the reconstruction of the $J/\psi\pi K$ invariant mass distributions for the reactions would show formation of the $K^*(4307)$ and the branching ratio determined for $B\to \pi K^*(4307)\to \pi K Z_c(3900)\to \pi K J/\psi\pi$ is $\sim 10^{-8}$. We hope that this calculation motivates the search of the $K^*(4307)$, formed as a consequence of the dynamics involved in the $KD\bar D^*$ system~\cite{Ma:2017ery,Ren:2018pcd}, by reconstructing the $J/\psi\pi K$ invariant mass distribution in $B\to J/\psi\pi\pi K$ reactions, for which experimental data are available. 

\section*{Acknowledgements}
The authors thanks Prof. Eulogio Oset for useful discussions. This work was partly supported by DFG and NSFC through funds provided to the Sino-German CRC 110 ``Symmetries and the Emergence of Structure in QCD'' (Grant No. TRR110), by the Funda\c c\~ao de Amparo \`a Pesquisa do Estado de S\~ao Paulo (FAPESP), processos n${}^\circ$ 2019/17149-3 and 2019/16924-3, by the Conselho Nacional de Desenvolvimento Cient\'ifico e Tecnol\'ogico (CNPq), grant  n${}^\circ$ 310759/2016-1 and 311524/2016-8, and by NSFC (grant n${}^\circ$ 11775099). 
\appendix
\section{Determination of the integrals in Eq.~(\ref{int})}\label{ap}
The first step for calculating the integrals in Eq.~(\ref{int}) consists of using the Passarino-Veltman decomposition of tensor integrals~\cite{Passarino:1978jh}, which exploits Lorentz covariance to write each of the integrals as a combination of the different Lorentz structures with some unknown coefficients. For example, the integral $I^{(1\mathcal{A})}_\alpha$ in Eq.~(\ref{int}) is a covariant tensor which can depend on the four-momenta $P$ and $p_\mathcal{A}$. In this way, we can write
\begin{align}
I^{(1\mathcal{A})}_\alpha=a^{(1\mathcal{A})}_1 P_\alpha+a^{(1\mathcal{A})}_2 p_{\mathcal{A}\,\alpha},~\label{I1}
\end{align}
where $a^{(1\mathcal{A})}_1$ and $a^{(1\mathcal{A})}_2$ are coefficients to be determined. Similarly, 
\begin{align}
I^{(2\mathcal{A})}_{\alpha\beta}&=a^{(2\mathcal{A})}_1g_{\alpha\beta}+a^{(2\mathcal{A})}_2P_\alpha P_\beta+a^{(2\mathcal{A})}_3(P_\alpha\, p_{\mathcal{A}\,\beta}+P_\beta\, p_{\mathcal{A}\,\alpha})+a^{(2\mathcal{A})}_4 p_{\mathcal{A}\,\alpha}\,p_{\mathcal{A}\,\beta},\nonumber\\
I^{(3\mathcal{A})}_\alpha&=a^{(3\mathcal{A})}_1 P_\alpha+a^{(3\mathcal{A})}_2 p_{\mathcal{A}\,\alpha},\nonumber\\
I^{(4\mathcal{A})}_{\alpha\beta}&=a^{(4\mathcal{A})}_1g_{\alpha\beta}+a^{(4\mathcal{A})}_2P_\alpha P_\beta+a^{(4\mathcal{A})}_3(P_\alpha\, p_{\mathcal{A}\,\beta}+P_\beta\, p_{\mathcal{A}\,\alpha})+a^{(4\mathcal{A})}_4 p_{\mathcal{A}\,\alpha}\,p_{\mathcal{A}\,\beta},\nonumber\\
I^{(5\mathcal{A})}_\alpha&=a^{(5\mathcal{A})}_1 P_\alpha+a^{(5\mathcal{A})}_2 p_{\mathcal{A}\,\alpha},\label{others}
\end{align}
where we have used the fact that the tensor integrals $I^{(2\mathcal{A})}_{\alpha\beta}$ and $I^{(4\mathcal{A})}_{\alpha\beta}$ are symmetric under the interchange $\alpha\leftrightarrow\beta$, as can be seen from the definition in Eq.~(\ref{int}). Contracting the integrals in Eqs.~(\ref{I1}) and (\ref{others}) with the different Lorentz structures appearing in their decomposition, we can get a system of equations which permits the determination of the unknown $a$-coefficients in terms of scalar integrals. For instance,
using Eq.~(\ref{I1}), we can write
\begin{align}
P\cdot I^{(1\mathcal{A})}&= a^{(1\mathcal{A})}_1 P^2+a^{(1\mathcal{A})}_2 P\cdot p_{\mathcal{A}},\nonumber\\
p_\mathcal{A}\cdot I^{(1\mathcal{A})}&=a^{(1\mathcal{A})}_1 P\cdot p_\mathcal{A}+a^{(1\mathcal{A})}_2 p^2_{\mathcal{A}}.\label{sys}
\end{align}
By solving the system of equations (\ref{sys}), we find
\begin{align}
a^{(1\mathcal{A})}_1&=-\frac{p^2_{\mathcal{A}}\, P\cdot I^{(1\mathcal{A})}-(P\cdot p_{\mathcal{A}})\, p_\mathcal{A}\cdot I^{(1\mathcal{A})}}{(P\cdot p_{\mathcal{A}})^2-P^2\,p^2_{\mathcal{A}}},\nonumber\\
a^{(1\mathcal{A})}_2&=-\frac{P^2\,p_\mathcal{A}\cdot I^{(1\mathcal{A})}-(P\cdot p_{\mathcal{A}})\, P\cdot I^{(1\mathcal{A})}}{(P\cdot p_{\mathcal{A}})^2-P^2\,p^2_{\mathcal{A}}},\label{a1}
\end{align}
and the whole problem reduces to determine the scalar integrals $P\cdot I^{(1\mathcal{A})}$ and $p_\mathcal{A}\cdot I^{(1\mathcal{A})}$, which, from Eq.~(\ref{int}), are given by
\begin{align}
P\cdot I^{(1\mathcal{A})}=\int\frac{d^4 q}{(2\pi)^4}\frac{P\cdot q}{\mathcal{D}(q,p_\mathcal{A})},\quad p_{\mathcal{A}}\cdot I^{(1\mathcal{A})}=\int\frac{d^4 q}{(2\pi)^4}\frac{p_\mathcal{A}\cdot q}{\mathcal{D}(q,p_\mathcal{A})}, \label{ppI1}
\end{align}
where $\mathcal{D}(q,p_\mathcal{A})$ can be found in Eq.~(\ref{den}). Similarly, the expression for the coefficients $a^{(3\mathcal{A})}_i$ and $a^{(5\mathcal{A})}_i$, with $i=1,2$, is analogous to the one found in Eq.~(\ref{a1}) for the $a^{(1\mathcal{A})}_i$ coefficients by replacing  $P\cdot I^{(1\mathcal{A})}\to P\cdot I^{(3\mathcal{A})}$, $p_{\mathcal{A}}\cdot I^{(1\mathcal{A})}\to p_{\mathcal{A}}\cdot I^{(3\mathcal{A})}$ in the former case, and 
$P\cdot I^{(1\mathcal{A})}\to P\cdot I^{(5\mathcal{A})}$, $p_{\mathcal{A}}\cdot I^{(1\mathcal{A})}\to p_{\mathcal{A}}\cdot I^{(5\mathcal{A})}$ in the latter, with
\begin{align}
P\cdot I^{(3\mathcal{A})}=\int\frac{d^4 q}{(2\pi)^4}\frac{q^2 (P\cdot q)}{\mathcal{D}(q,p_\mathcal{A})},\quad p_{\mathcal{A}}\cdot I^{(3\mathcal{A})}=\int\frac{d^4 q}{(2\pi)^4}\frac{q^2(p_\mathcal{A}\cdot q)}{\mathcal{D}(q,p_\mathcal{A})},\nonumber\\
P\cdot I^{(5\mathcal{A})}=\int\frac{d^4 q}{(2\pi)^4}\frac{q^4(P\cdot q)}{\mathcal{D}(q,p_\mathcal{A})},\quad p_{\mathcal{A}}\cdot I^{(5\mathcal{A})}=\int\frac{d^4 q}{(2\pi)^4}\frac{q^4(p_\mathcal{A}\cdot q)}{\mathcal{D}(q,p_\mathcal{A})}.\label{ppI35}
\end{align}
Proceeding in the same way as in case of the $a^{(1\mathcal{A})}_i$ coefficients (see Eq.~(\ref{a1})), we have 
\begin{align}
a^{(2\mathcal{A})}_1&=\frac{1}{2\left[(P\cdot p_\mathcal{A})^2- P^2\,\,p^2_{\mathcal{A}}\right]}
\Bigg[P^2\left(p_\mathcal{A}\cdot p_\mathcal{A}\cdot I^{(2\mathcal{A})}-p^2_\mathcal{A}\,g\cdot I^{(2\mathcal{A})}\right)+(P\cdot p_\mathcal{A})^2\, g\cdot I^{(2\mathcal{A})}\nonumber\\
&\quad-2(P\cdot p_\mathcal{A})\, P\cdot p_\mathcal{A}\cdot I^{(2\mathcal{A})}+p^2_\mathcal{A}\,P\cdot P\cdot I^{(2\mathcal{A})}\Bigg],\nonumber\\
a^{(2\mathcal{A})}_2&=\frac{1}{2\left[(P\cdot p_\mathcal{A})^2- P^2\,\,p^2_{\mathcal{A}}\right]^2}\Bigg[p^2_\mathcal{A}\left\{P^2\left(p_\mathcal{A}\cdot p_\mathcal{A}\cdot I^{(2\mathcal{A})}-p^2_\mathcal{A}\,g\cdot I^{(2\mathcal{A})}\right)+3p^2_\mathcal{A}\,P\cdot P\cdot I^{(2\mathcal{A})}\right\}\nonumber\\
&\quad+(P\cdot p_\mathcal{A})^2\left\{p^2_\mathcal{A}\,g\cdot I^{(2\mathcal{A})}+2\,p_\mathcal{A}\cdot p_\mathcal{A}\cdot I^{(2\mathcal{A})}\right\}-6 p^2_\mathcal{A}\,(P\cdot p_\mathcal{A})\,P\cdot p_\mathcal{A}\cdot I^{(2\mathcal{A})}\Bigg],\nonumber\\
a^{(2\mathcal{A})}_3&=\frac{1}{2\left[(P\cdot p_\mathcal{A})^2- P^2\,\,p^2_{\mathcal{A}}\right]^2}\Bigg[P^2\Big\{(P\cdot p_\mathcal{A})\left(p^2_\mathcal{A}\,\,g\cdot I^{(2\mathcal{A})}-3\,p_\mathcal{A}\cdot p_\mathcal{A}\cdot I^{(2\mathcal{A})}\right)\nonumber\\
&\quad+2 p^2_\mathcal{A}\,P\cdot p_\mathcal{A}\cdot I^{(2\mathcal{A})}\Big\}-(P\cdot p_\mathcal{A})\Big\{(P\cdot p_\mathcal{A})^2\,\,g\cdot I^{(2\mathcal{A})}-4(P\cdot p_\mathcal{A})\,P\cdot p_\mathcal{A}\cdot I^{(2\mathcal{A})}\nonumber\\
&\quad+3 p^2_\mathcal{A}\,\, P\cdot P\cdot I^{(2\mathcal{A})}\Big\}\Bigg],\nonumber\\
a^{(2\mathcal{A})}_4&=\frac{1}{2\left[(P\cdot p_\mathcal{A})^2- P^2\,\,p^2_{\mathcal{A}}\right]^2}\Bigg[P^4\Big\{3\,p_\mathcal{A}\cdot p_\mathcal{A}\cdot I^{(2\mathcal{A})}-p^2_\mathcal{A}\,\, g\cdot I^{(2\mathcal{A})}\Big\}\nonumber\\
&\quad+P^2\Big\{(P\cdot p_\mathcal{A})^2\,\, g\cdot I^{(2\mathcal{A})}-6\,(P\cdot p_\mathcal{A})\,\, P\cdot p_\mathcal{A}\cdot I^{(2\mathcal{A})}+p^2_\mathcal{A}\,\, P\cdot P\cdot I^{(2\mathcal{A})}\Big\}\nonumber\\
&\quad+2\,(P\cdot p_\mathcal{A})^2\,\, P\cdot P\cdot I^{(2\mathcal{A})}\Bigg].\label{a24}
\end{align}
where
\begin{align}
p_\mathcal{A}\cdot p_\mathcal{A}\cdot I^{(2\mathcal{A})}&=\int\frac{d^4 q}{(2\pi)^4}\frac{(p_\mathcal{A}\cdot q)(p_\mathcal{A}\cdot q)}{\mathcal{D}(q,p_\mathcal{A})},\quad 
P\cdot p_\mathcal{A}\cdot I^{(2\mathcal{A})}=\int\frac{d^4 q}{(2\pi)^4}\frac{(P\cdot q)(p_\mathcal{A}\cdot q)}{\mathcal{D}(q,p_\mathcal{A})},\nonumber\\
P\cdot P\cdot I^{(2\mathcal{A})}&=\int\frac{d^4 q}{(2\pi)^4}\frac{(P\cdot q)(P\cdot q)}{\mathcal{D}(q,p_\mathcal{A})},\quad g\cdot I^{(2\mathcal{A})}=\int\frac{d^4 q}{(2\pi)^4}\frac{q^2}{\mathcal{D}(q,p_\mathcal{A})}.\label{ppI2}
\end{align}

Analogously, for the coefficients $a^{(4\mathcal{A})}_i$, $i=1,2,\dots,4$, we can simply replace in Eqs.~(\ref{a24}) the scalar integrals $p_\mathcal{A}\cdot p_\mathcal{A}\cdot I^{(2\mathcal{A})}$, $P\cdot p_\mathcal{A}\cdot I^{(2\mathcal{A})}$, $P\cdot P\cdot I^{(2\mathcal{A})}$ and $g\cdot I^{(2\mathcal{A})}$ by $p_\mathcal{A}\cdot p_\mathcal{A}\cdot I^{(4\mathcal{A})}$, $P\cdot p_\mathcal{A}\cdot I^{(4\mathcal{A})}$, $P\cdot P\cdot I^{(4\mathcal{A})}$ and $g\cdot I^{(4\mathcal{A})}$, respectively, with
\begin{align}
p_\mathcal{A}\cdot p_\mathcal{A}\cdot I^{(4\mathcal{A})}&=\int\frac{d^4 q}{(2\pi)^4}\frac{q^2(p_\mathcal{A}\cdot q)(p_\mathcal{A}\cdot q)}{\mathcal{D}(q,p_\mathcal{A})},\quad 
P\cdot p_\mathcal{A}\cdot I^{(4\mathcal{A})}=\int\frac{d^4 q}{(2\pi)^4}\frac{q^2(P\cdot q)(p_\mathcal{A}\cdot q)}{\mathcal{D}(q,p_\mathcal{A})},\nonumber\\
P\cdot P\cdot I^{(4\mathcal{A})}&=\int\frac{d^4 q}{(2\pi)^4}\frac{q^2(P\cdot q)(P\cdot q)}{\mathcal{D}(q,p_\mathcal{A})},\quad g\cdot I^{(4\mathcal{A})}=\int\frac{d^4 q}{(2\pi)^4}\frac{q^4}{\mathcal{D}(q,p_\mathcal{A})}\label{ppI4}
\end{align}

The next step consists in calculating the scalar integrals in Eqs.~(\ref{ppI1}),~(\ref{ppI35}),~(\ref{ppI2}),~(\ref{ppI4}), which we do in the rest frame of the decaying particle, i.e., $P^2=m^2_{B^+}$ and $\vec{P}=\vec{0}$. To do this, it is convenient to realize that these integrals can be considered as particular cases of other more general integrals. For instance, if we define $\mathcal{I}^{(1\mathcal{A})}(a,b)$ as
\begin{align}
\mathcal{I}^{(1\mathcal{A})}(a,b)=\int\frac{dq^0}{(2\pi)}\int\frac{d^3 q}{(2\pi)^3}\frac{a q^0+b|\vec{q}| cos\theta}{\mathcal{D}(q,p_\mathcal{A})},\label{Im1}
\end{align}
we can write
\begin{align}
P\cdot I^{(1\mathcal{A})}=\mathcal{I}^{(1\mathcal{A})}(P^0,0),\quad p_\mathcal{A}\cdot I^{(1\mathcal{A})}=\mathcal{I}^{(1\mathcal{A})}(p^0_\mathcal{A},-|\vec{p}_\mathcal{A}|).
\end{align}
Similarly, by defining
\begin{align}
\mathcal{I}^{(2\mathcal{A})}(a,b,c,d)&=\int\frac{dq^0}{(2\pi)}\int\frac{d^3 q}{(2\pi)^3}\frac{a q^{0\,2}+b|\vec{q}\,|^{2} \text{cos}^2\theta+c q^0 |\vec{q}\,|\text{cos}\theta+d|\vec{q}\,|^2}{\mathcal{D}(q,p_\mathcal{A})},\nonumber\\
\mathcal{I}^{(3\mathcal{A})}(a,b,c,d)&=\int\frac{dq^0}{(2\pi)}\int\frac{d^3 q}{(2\pi)^3}\frac{a q^{0\,3}+b|\vec{q}\,|q^{0\,2} \text{cos}\theta+c q^0 |\vec{q}\,|^2+d|\vec{q}\,|^3\text{cos}\theta}{\mathcal{D}(q,p_\mathcal{A})},\nonumber\\
\mathcal{I}^{(4\mathcal{A})}(a,b,c,d,e,f,g)&=\int\frac{dq^0}{(2\pi)}\int\frac{d^3 q}{(2\pi)^3}\frac{1}{\mathcal{D}(q,p_\mathcal{A})}\Big[a q^{0\,4}+b|\vec{q}\,|^2 q^{0\,2} \text{cos}^2\theta+c q^{0\,3} |\vec{q}\,|\text{cos}\theta\nonumber\\
&\quad+d q^{0\,2}|\vec{q}\,|^2+e|\vec{q}\,|^4\text{cos}^2\theta+f q^0 |\vec{q}\,|^3\text{cos}\theta+g|\vec{q}\,|^4\Big],\nonumber\\
\mathcal{I}^{(5\mathcal{A})}(a,b,c,d,e,f)&=\int\frac{dq^0}{(2\pi)}\int\frac{d^3 q}{(2\pi)^3}\frac{1}{\mathcal{D}(q,p_\mathcal{A})}\Big[a q^{0\,5}+b q^{0\,4}|\vec{q}\,| \text{cos}\theta+c q^{0\,3} |\vec{q}\,|^2\nonumber\\
&\quad+d q^{0\,2}|\vec{q}\,|^3\text{cos}\theta+e q^0 |\vec{q}\,|^4+f  |\vec{q}\,|^5\text{cos}\theta\Big],\label{Imothers}
\end{align}
we have that
\begin{align}
g\cdot I^{(2\mathcal{A})}&=\mathcal{I}^{(2\mathcal{A})}(1,0,0,-1),\quad P\cdot P\cdot I^{(2\mathcal{A})}=\mathcal{I}^{(2\mathcal{A})}(P^{0\,2},0,0,0),\nonumber\\
P\cdot p_\mathcal{A}\cdot I^{(2\mathcal{A})}&=\mathcal{I}^{(2\mathcal{A})}(P^0\,p^0_\mathcal{A},0,-P^0\,|\vec{p}_\mathcal{A}|,0),\nonumber\\
p_\mathcal{A}\cdot p_\mathcal{A}\cdot I^{(2\mathcal{A})}&=\mathcal{I}^{(2\mathcal{A})}(p^{0\,2}_\mathcal{A},|\vec{p}_\mathcal{A}\,|^2,-2p^0_\mathcal{A}|\vec{p}_\mathcal{A}|,0),\nonumber\\
P\cdot I^{(3\mathcal{A})}&=\mathcal{I}^{(3\mathcal{A})}(P^0,0,-P^0,0),\quad p_\mathcal{A}\cdot I^{(3\mathcal{A})}=\mathcal{I}^{(3\mathcal{A})}(p^0_\mathcal{A},-|\vec{p}_\mathcal{A}|,-p^0_\mathcal{A},|\vec{p}_\mathcal{A}|),\nonumber\\
g\cdot I^{(4\mathcal{A})}&=\mathcal{I}^{(4\mathcal{A})}(1,0,0,-2,0,0,1),\quad P\cdot P\cdot I^{(4\mathcal{A})}=\mathcal{I}^{(4\mathcal{A})}(P^{0\,2},0,0,-P^{0\,2},0,0,0),\nonumber\\
P\cdot p_\mathcal{A}\cdot I^{(4\mathcal{A})}&=\mathcal{I}^{(4\mathcal{A})}(P^0 p^0_\mathcal{A},0,-P^0|\vec{p}_\mathcal{A}|,-P^0 p^0_\mathcal{A},0,P^0|\vec{p}_\mathcal{A}|,0),\nonumber\\
p_\mathcal{A}\cdot p_\mathcal{A}\cdot I^{(4\mathcal{A})}&=\mathcal{I}^{(4\mathcal{A})}(p^{0\,2}_\mathcal{A},|\vec{p}_\mathcal{A}|^2,-2p^0_\mathcal{A}|\vec{p}_\mathcal{A}|,-p^{0\,2}_\mathcal{A},-|\vec{p}_\mathcal{A}|^2,2p^0_\mathcal{A}|\vec{p}_\mathcal{A}|,0),\nonumber\\
P\cdot I^{(5\mathcal{A})}&=\mathcal{I}^{(5\mathcal{A})}(P^0,0,-2P^0,0,P^0,0),\nonumber\\
p_\mathcal{A}\cdot I^{(5\mathcal{A})}&=\mathcal{I}^{(5\mathcal{A})}(p^0_\mathcal{A},-|\vec{p}_\mathcal{A}|,-2p^0_\mathcal{A},2|\vec{p}_\mathcal{A}|,p^0_\mathcal{A},-|\vec{p}_\mathcal{A}|).
\end{align}
We also define
\begin{align}
\mathcal{I}^{(0\mathcal{A})}=\int\frac{dq^0}{(2\pi)}\int\frac{d^3 q}{(2\pi)^3}\frac{1}{\mathcal{D}(q,p_\mathcal{A})},\label{I0A}
\end{align}
which coincides with the $I^{(0\mathcal{A})}$ integral in Eq.~(\ref{int}).

After this, it is convenient to separate the temporal part in the denominator $\mathcal{D}(q,p_\mathcal{A})$ and write it as 
\begin{align}
\mathcal{D}(q,p_\mathcal{A})&=[(P^0-q^0)^2-\omega^2_{J/\psi}(\vec{q})+i\epsilon][q^{0\,2}-\omega^2_K(\vec{q})+i\epsilon]\nonumber\\
&\quad\times[(P^0-p^0_\mathcal{A}-q^0)^2-\omega_{Z(\mathcal{A})}(\vec{p}_\mathcal{A}+\vec{q})+i\epsilon],
\end{align}
where
\begin{align}
\omega_{J/\psi}(\vec{q})=\sqrt{\vec{q}^{\,\,2}+m^2_{J/\psi}},\quad \omega_K(\vec{q})=\sqrt{\vec{q}^{\,\,2}+m^2_{K^+}},\nonumber\\
\omega_{Z(\mathcal{A})}(\vec{p}_\mathcal{A}+\vec{q})=\sqrt{\vec{p}^{\,\,2}_\mathcal{A}+\vec{q}^{\,\,2}+2|\vec{p}_\mathcal{A}||\vec{q}\,|\text{cos}\theta+M^2_{Z(\mathcal{A})}},
\end{align}
with $\theta$ being the angle between $\vec{p}_\mathcal{A}$ and $\vec{q}$. By doing this, we find the following poles on the $q^0$ variable
\begin{align}
q^0_1=P^0+\omega_{J/\psi}-i\epsilon,&\quad q^0_2=P^0-\omega_{J/\psi}+i\epsilon,\nonumber\\
q^0_3=\omega_K-i\epsilon,&\quad q^0_4=-\omega_K+i\epsilon,\\
q^0_5=-p^0_3+P^0+\omega_{Z(\mathcal{A})}-i\epsilon,&\quad q^0_6=-p^0_3+P^0-\omega_{Z(\mathcal{A})}+i\epsilon.
\end{align}
By using Cauchy's theorem,  the integration on the $q^0$ variable in Eqs.~(\ref{Im1}),~(\ref{Imothers}) and (\ref{I0A}) can be performed analytically by determining the residues of the respective integrands at each of the poles inside a closed contour, for example, a semicircle in the lower $q^0$ complex plane, i.e., $\text{Im}\{q^0\}<0$, involving, in this case, the poles $q^0_1$, $q^0_3$ and $q^0_5$. After doing this, the integrals in Eqs.~(\ref{Im1}),~(\ref{Imothers}) and (\ref{I0A}) can be written as
\begin{align}
\mathcal{I}^{(k\mathcal{A})}(a,b,\dots)=\int\frac{d^3q}{(2\pi)^3}\frac{\mathcal{N}^{(k\mathcal{A})}(a,b,\dots)}{\mathcal{D}(\vec{q},p^0_\mathcal{A},\vec{p}_\mathcal{A})},\label{ND}
\end{align}
with $k=0,1,2,\dots 4$ and
\begin{align}
\mathcal{D}(\vec{q},p^0_\mathcal{A},\vec{p}_\mathcal{A})&=2\omega_{J/\psi}(\vec{q}\,)\omega_K(\vec{q}\,)\omega_{Z(\mathcal{A})}(\vec{p}_\mathcal{A}+\vec{q}\,)\Big[P^0+\omega_{J/\psi}(\vec{q}\,)+\omega_K(\vec{q}\,)\Big]\nonumber\\
&\quad\times\Big[p^0_\mathcal{A}+\omega_{J/\psi}(\vec{q}\,)+\omega_{Z(\mathcal{A})}(\vec{p}_\mathcal{A}+\vec{q}\,)\Big]\Big[P^0-\omega_K(\vec{q}\,)-\omega_{J/\psi}(\vec{q}\,)+i\epsilon\Big]\nonumber\\
&\quad\times\Big[P^0-p^0_\mathcal{A}-\omega_K(\vec{q}\,)-\omega_{Z(\mathcal{A})}(\vec{p}_\mathcal{A}+\vec{q}\,)+i\frac{\Gamma_{Z(\mathcal{A})}}{2}\Big]\nonumber\\
&\quad\times\Big[p^0_\mathcal{A}-\omega_{J/\psi}(\vec{q}\,)-\omega_{Z(\mathcal{A})}(\vec{p}_\mathcal{A}+\vec{q}\,)+i\frac{\Gamma_{Z(\mathcal{A})}}{2}\Big]\nonumber\\
&\quad\times\Big[p^0_\mathcal{A}-P^0-\omega_K(\vec{q}\,)-\omega_{Z(\mathcal{A})}(\vec{p}_\mathcal{A}+\vec{q}\,)+i\frac{\Gamma_{Z(\mathcal{A})}}{2}\Big].\label{D2}
\end{align}
In Eq.~(\ref{D2}), a width $\Gamma_{Z(\mathcal{A})}$ of 28 MeV~\cite{Tanabashi:2018oca} has been considered for the $Z_c(3900)$ present in the triangular loops. The $\mathcal{N}^{(k\mathcal{A})}$ numerators in Eq.~(\ref{ND}) are given by
\begin{align}
\mathcal{N}^{(0\mathcal{A})}&=-P^{0\,2}\omega_K\omega_{J/\psi+Z(\mathcal{A})}+2 P^0 p^0_\mathcal{A}\omega_K\omega_{Z(\mathcal{A})}+\omega_{J/\psi+K}\Big[\omega_{J/\psi+Z(\mathcal{A})}\nonumber\\
&\quad\times \omega_{K+Z(\mathcal{A})}\omega_{J/\psi+K+Z(\mathcal{A})}-p^{0\,2}_\mathcal{A}\,\omega_{Z(\mathcal{A})}\Big],\nonumber\\
\mathcal{N}^{(1\mathcal{A})}(a,b)&=a\,\omega_K f_1+\text{cos}\theta\,b|\vec{q}\,| \mathcal{N}^{(0\mathcal{A})},\nonumber\\
\mathcal{N}^{(2\mathcal{A})}(a,b,c,d)&=a\,\omega_K f_2+|\vec{q}\,|\left[(\text{cos}^2\theta\,b+d)|\vec{q}\,|\mathcal{N}^{(0\mathcal{A})}+\text{cos}\theta\,c\,\omega_K f_1\right],\nonumber\\
\mathcal{N}^{(3\mathcal{A})}(a,b,c,d)&=a\,\omega_K f_3+|\vec{q}\,|\left[\text{cos}\theta\left\{b\,\omega_K f_2+d|\vec{q}\,|^2 \mathcal{N}^{(0\mathcal{A})}\right\}+c|\vec{q}\,|\omega_K f_1\right]\nonumber\\
\mathcal{N}^{(4\mathcal{A})}(a,b,c,d,e,f,g)&=a\,\omega_K f_4+|\vec{q}\,|\Big[\text{cos}\theta\Big\{|\vec{q}\,|\Big(\text{cos}\theta [b\,\omega_K f_2+e|\vec{q}\,|^2 \mathcal{N}^{(0\mathcal{A})}]+f\,|\vec{q}\,|\omega_K f_1\Big)\nonumber\\
&\quad+c\,\omega_K f_3\Big\}+d\,|\vec{q}\,|\omega_K f_2+g\,|\vec{q}\,|^3\mathcal{N}^{(0\mathcal{A})}\Big],\nonumber\\
\mathcal{N}^{(5\mathcal{A})}(a,b,c,d,e,f)&=a\,\omega_K f_5+|\vec{q}\,|\Big[\omega_K|\vec{q}\,|\Big\{f_1|\vec{q}\,|^2\,e+c\,f_3\Big\}\nonumber\\
&\quad+\Big\{f\,\mathcal{N}^{(0\mathcal{A})}|\vec{q}\,|^4+\omega_K(d\,f_2|\vec{q}\,|^2+b\,f_4)\Big\}\text{cos}\theta\Big],\label{Ns}
\end{align}
where we have omitted the explicit dependence of the functions $f_i$, $i=1,2,\dots,5$ and $\omega_K$, $\omega_{Z(\mathcal{A})}$, $\omega_{J/\psi}$ [see Eq.~(\ref{D2})] with $\vec{q}$, $\vec{p}_\mathcal{A}$ and $p^0_\mathcal{A}$ for simplicity. In particular, by introducing
\begin{align}
\omega_{J/\psi+K}&=\omega_{J/\psi}+\omega_K,\quad\omega_{J/\psi+Z(\mathcal{A})}=\omega_{J/\psi}+\omega_{Z(\mathcal{A})},\nonumber\\
\omega_{K+Z(\mathcal{A})}&=\omega_K+\omega_{Z(\mathcal{A})},\quad\omega_{J/\psi+K+Z(\mathcal{A})}=\omega_{J/\psi}+\omega_K+\omega_{Z(\mathcal{A})},
\end{align}
the $f_i$ functions in Eq.~(\ref{Ns}) correspond to
\begin{align}
f_1(\vec{q},p^0_\mathcal{A},\vec{p}_\mathcal{A})&=P^{0\,2} p^0_\mathcal{A}\omega_{Z(\mathcal{A})}+\omega_{J/\psi+Z(\mathcal{A})}P^0\Big(-P^{0\,2}+P^0 p^0_\mathcal{A}
+\omega_{J/\psi}\left[\omega_{J/\psi+Z(\mathcal{A})}+2\omega_K\right]\nonumber\\
&\quad+\omega_{K+Z(\mathcal{A})}^2\Big)-P^0 p^{0\,2}_\mathcal{A}\omega_{Z(\mathcal{A})}-p^0_\mathcal{A}\omega_{J/\psi}\omega_{J/\psi+K}[\omega_{J/\psi+K}+2\omega_{Z(\mathcal{A})}],\nonumber\\
f_2(\vec{q},p^0_\mathcal{A},\vec{p}_\mathcal{A})&=\omega_{J/\psi+Z(\mathcal{A})}\left[-P^{0\,2}+2P^0 p^0_\mathcal{A}-p^{0\,2}_{\mathcal{A}}+\omega_{J/\psi+K}^2+2\omega_K\omega_{Z(\mathcal{A})}+\omega^2_{Z(\mathcal{A})}\right]P^{0\,2}\nonumber\\
&\quad-2P^0 p^0_\mathcal{A}\omega_{J/\psi}\Big[\omega_{J/\psi+K}\omega_{J/\psi+K+Z(\mathcal{A})}+\omega_{Z(\mathcal{A})}\omega_K\Big]\nonumber\\
&\quad+\omega_{J/\psi}\omega_{J/\psi+K}\Big[\omega_{J/\psi+Z(\mathcal{A})}\left\{p^{0\,2}_\mathcal{A}-\omega_{Z(\mathcal{A})}\omega_{K+Z(\mathcal{A})}\right\}+p^{0\,2}_\mathcal{A}\omega_K\Big],\nonumber\\
f_3(\vec{q},p^0_\mathcal{A},\vec{p}_\mathcal{A})&=-P^{0\,5}\omega_{J/\psi+Z(\mathcal{A})}+P^{0\,4}p^0_\mathcal{A}[2\omega_{J/\psi+Z(\mathcal{A})}+\omega_{J/\psi}]+P^{0\,3}\Big(\omega_{J/\psi+Z(\mathcal{A})}\nonumber\\
&\quad\times\Big[\omega^2_{J/\psi+K}+2\omega_K\omega_{Z(\mathcal{A})}+\omega^2_{Z(\mathcal{A})}\Big]-p^{0\,2}_\mathcal{A}\Big[\omega_{J/\psi+Z(\mathcal{A})}+2\omega_{J/\psi}\Big]\Big)+P^{0\,2}p^0_\mathcal{A}\omega_{J/\psi}\nonumber\\
&\quad\times\Big(p^{0\,2}_\mathcal{A}-2\omega_{Z(\mathcal{A})}[\omega_{J/\psi+K}+2\omega_K]-3\omega_{J/\psi+K}^2-\omega^2_{Z(\mathcal{A})}\Big)+P^0\omega_{J/\psi}\nonumber\\
&\quad\times\Big(p^{0\,2}_\mathcal{A}\Big[\omega_{Z(\mathcal{A})}\left\{\omega_{J/\psi+K}+\omega_K\right\}+3\omega_{J/\psi+K}^2\Big]-\omega_{Z(\mathcal{A})}\omega_{J/\psi+Z(\mathcal{A})}\nonumber\\
&\quad\times\Big[2\omega_{J/\psi+K}\omega_{K+Z(\mathcal{A})}+\omega^2_K\Big]\Big)+p^0_\mathcal{A}\omega_{J/\psi}\omega_{J/\psi+K}\nonumber\\
&\quad\times\Big(\omega_{Z(\mathcal{A})}\left[\omega_{Z(\mathcal{A})}\omega_{J/\psi+K}+2\omega_{J/\psi}\omega_K\right]-p^{0\,2}_\mathcal{A}\omega_{J/\psi+K}\Big),\nonumber
\end{align}
\begin{align}
f_4(\vec{q},p^0_\mathcal{A},\vec{p}_\mathcal{A})&=-P^{0\,6}\omega_{J/\psi+Z(\mathcal{A})}+2P^{0\,5}p^0_\mathcal{A}\left(\omega_{J/\psi+Z(\mathcal{A})}+\omega_{J/\psi}\right)+P^{0\,4}\Big(\omega_{J/\psi+Z(\mathcal{A})}\nonumber\\
&\quad\times\Big[\omega_{Z(\mathcal{A})}\{\omega_{J/\psi+K}+\omega_K\}+\omega_{J/\psi+K}^2+\omega^2_{Z(\mathcal{A})}\Big]-p^{0\,2}_\mathcal{A}\Big[\omega_{J/\psi+Z(\mathcal{A})}+5\omega_{J/\psi}\Big]\Big)\nonumber\\
&\quad+4 P^{0\,3}p^0_\mathcal{A}\omega_{J/\psi}\Big(p^{0\,2}_\mathcal{A}-\omega_{Z(\mathcal{A})}[\omega_{J/\psi+K}+\omega_{K+Z(\mathcal{A})}]-\omega_{J/\psi+K}^2\Big)\nonumber\\
&\quad-P^{0\,2}\omega_{J/\psi}\Big(p^{0\,4}_\mathcal{A}-2p^{0\,2}_\mathcal{A}\Big[\omega_{Z(\mathcal{A})}\{\omega_{J/\psi+K}+\omega_{K+Z(\mathcal{A})}\}+3\omega_{J/\psi+K}^2\Big]\nonumber\\
&\quad+\omega_{Z(\mathcal{A})}\omega_{J/\psi+Z(\mathcal{A})}\Big[\omega_{J/\psi}\left\{\omega_{J/\psi+K}+3\omega_{K}\right\}+2\omega_K\left\{2\omega_{K+Z(\mathcal{A})}+\omega_K\right\}+\omega^2_{Z(\mathcal{A})}\Big]\Big)\nonumber\\
&\quad+2P^0 p^0_\mathcal{A}\omega_{J/\psi}\Big(\omega_{Z(\mathcal{A})}\Big[2\omega_{Z(\mathcal{A})}\omega_{J/\psi+K}^2+\omega_{J/\psi}\left\{\omega_{J/\psi+K}+\omega_K\right\}^2\Big]-2p^{0\,2}_\mathcal{A}\omega^2_{J/\psi+K}\Big)\nonumber\\
&\quad+\omega_{J/\psi}\omega_{J/\psi+K}\Big(p^{0\,4}_3\omega_{J/\psi+K}-p^{0\,2}_\mathcal{A}\omega_{Z(\mathcal{A})}\Big[\omega^2_{J/\psi+K}+2\omega_{Z(\mathcal{A})}\omega_{J/\psi+K}+\omega_{J/\psi}\omega_K\Big]\nonumber\\
&\quad+\omega_{Z(\mathcal{A})}\omega_{J/\psi+Z(\mathcal{A})}\omega_{K+Z(\mathcal{A})}\Big[\omega_{Z(\mathcal{A})}\omega_{J/\psi+K}+\omega_{J/\psi}\omega_K\Big]\Big),\nonumber
\end{align}
\begin{align}
f_5(\vec{q},p^0_\mathcal{A},\vec{p}_\mathcal{A})&=-\omega_{J/\psi+Z(\mathcal{A})}P^{0\,7}+p^0_\mathcal{A}\left(2\omega_{J/\psi+Z(\mathcal{A})}+3\omega_{J/\psi}\right)P^{0\,6}+\Big(\omega_{J/\psi+Z(\mathcal{A})}\Big[\omega^2_{J/\psi}+\left\{2\omega_K+3\omega_{Z(\mathcal{A})}\right\}\nonumber\\
&\quad\times\omega_{J/\psi}+\omega^2_{K+Z(\mathcal{A})}\Big]-p^{0\,2}_\mathcal{A}\left[\omega_{J/\psi+Z(\mathcal{A})}+9\omega_{J/\psi}\right]\Big)P^{0\,5}+\omega_{J/\psi}\Big(10p^{0\,3}_\mathcal{A}-\omega_{Z(\mathcal{A})}p^{0\,2}_\mathcal{A}\nonumber\\
&\quad-5\left[\omega_{J/\psi+K}\{\omega_{J/\psi+K}+2\omega_{Z(\mathcal{A})}\}+2\omega^2_{Z(\mathcal{A})}\right]p^0_\mathcal{A}+\omega_{Z(\mathcal{A})}\omega_{J/\psi+Z(\mathcal{A})}^2\Big)P^{0\,4}\nonumber\\
&\quad-\omega_{J/\psi}\Big(5 p^{0\,4}_\mathcal{A}-2\omega_{Z(\mathcal{A})}p^{0\,3}_\mathcal{A}+2\Big[-\omega_{J/\psi+K}\left\{5\omega_{J/\psi+K}+4\omega_{Z(\mathcal{A})}\right\}-4\omega^2_{Z(\mathcal{A})}\Big]p^{0\,2}_\mathcal{A}\nonumber\\
&\quad+2\omega_{Z(\mathcal{A})}\omega^2_{J/\psi+Z(\mathcal{A})}p^0_\mathcal{A}+\omega_{Z(\mathcal{A})}\omega_{J/\psi+Z(\mathcal{A})}\Big[\omega_{J/\psi}\{3\omega_{J/\psi}+\omega_{K+Z(\mathcal{A})}+7\omega_K\}\nonumber\\
&\quad+10 \omega^2_K+\omega_{Z(\mathcal{A})}\{3\omega_{K+Z(\mathcal{A})}+5\omega_K\}\Big]\Big)P^{0\,3}-\omega_{J/\psi}\Big(-p^{0\,5}_\mathcal{A}+\omega_{Z(\mathcal{A})}p^{0\,4}_\mathcal{A}\nonumber\\
&\quad+2\Big[\omega_{J/\psi+K}\left\{5\omega_{J/\psi+K}+\omega_{Z(\mathcal{A})}\right\}+\omega^2_{Z(\mathcal{A})}\Big]p^{0\,3}_\mathcal{A}-2\omega_{Z(\mathcal{A})}\Big[\omega_{J/\psi}\left\{\omega_{J/\psi}+\omega_{K+Z(\mathcal{A})}\right\}\nonumber\\
&\quad+\omega^2_{K+Z(\mathcal{A})}-\omega_K\omega_{Z(\mathcal{A})}\Big]p^{0\,2}_\mathcal{A}-\omega_{Z(\mathcal{A})}\Big[\omega^2_{Z(\mathcal{A})}\left\{\omega_{Z(\mathcal{A})}+2\omega_{J/\psi+K}\right\}\nonumber\\
&\quad+2\{5\omega^2_{J/\psi}+12\omega_K\omega_{J/\psi}+5\omega^2_K\}\omega_{Z(\mathcal{A})}+2\omega_{J/\psi}\{4\omega^2_{J/\psi}+11\omega_K\omega_{J/\psi}+10\omega^2_K\}\Big]p^0_\mathcal{A}\nonumber\\
&\quad+\omega_{Z(\mathcal{A})}\omega^2_{J/\psi+Z(\mathcal{A})}\Big[\omega^2_{J/\psi+K}+\omega^2_{K+Z(\mathcal{A})}\Big]\Big) P^{0\,2}+\omega_{J/\psi}\Big(\omega_{J/\psi+K}^2 \Big[5p^{0\,4}_\mathcal{A}\nonumber\\
&\quad-2\omega_{Z(\mathcal{A})}p^{0\,3}_\mathcal{A}\Big]-\omega_{Z(\mathcal{A})}\Big[7\omega^3_{J/\psi}+20\omega_K\omega^2_{J/\psi}+18 \omega^2_K\omega_{J/\psi}+4\omega^3_K+8\omega_{J/\psi+K}^2\omega_{Z(\mathcal{A})}\Big]p^{0\,2}_\mathcal{A}\nonumber\\
&\quad+2\omega_{J/\psi+K}^2\omega_{Z(\mathcal{A})}\omega_{J/\psi+Z(\mathcal{A})}^2p^0_\mathcal{A}+\omega_{Z(\mathcal{A})}\omega_{J/\psi+Z(\mathcal{A})}\Big[\omega_{Z(\mathcal{A})}\{3\omega_{K+Z(\mathcal{A})}+\omega_K\}\omega^2_K\nonumber\\
&\quad+\omega_{J/\psi}\{\omega_{K+Z(\mathcal{A})}+\omega_{Z(\mathcal{A})}\}\{3\omega_{K+Z(\mathcal{A})}+\omega_K\}\omega_K+3\omega^2_{J/\psi}\omega^2_{K+Z(\mathcal{A})}\Big]\Big)P^0\nonumber\\
&\quad+\omega_{J/\psi}\omega_{J/\psi+K}\Big(\omega_{J/\psi+K}\Big[-p^{0\,5}_\mathcal{A}+p^{0\,4}_\mathcal{A}\omega_{Z(\mathcal{A})}+2\omega_{Z(\mathcal{A})}\omega_{J/\psi+K+Z(\mathcal{A})}p^{0\,3}_\mathcal{A}\nonumber\\
&\quad-\omega_{Z(\mathcal{A})}\Big\{\omega_{J/\psi}\left(\omega_{J/\psi+Z(\mathcal{A})}+\omega_{Z(\mathcal{A})}\right)+\omega^2_K+2\omega_{Z(\mathcal{A})}\omega_{K+Z(\mathcal{A})}\Big\}p^{0\,2}_\mathcal{A}\Big]\nonumber\\
&\quad+\omega_{Z(\mathcal{A})}\Big[-\omega_{Z(\mathcal{A})}\Big\{\omega_{K+Z(\mathcal{A})}+\omega_K\Big\}\Big\{\omega_K\omega_{Z(\mathcal{A})}+\omega_{J/\psi}(\omega_{K+Z(\mathcal{A})}+\omega_K)\Big\}\nonumber\\
&\quad-2\omega^2_{J/\psi}\omega^2_{K+Z(\mathcal{A})}\Big]p^0_\mathcal{A}+\omega_{J/\psi+K}\omega_{Z(\mathcal{A})}\omega^2_{J/\psi+Z(\mathcal{A})}\omega^2_{K+Z(\mathcal{A})}\Big).
\end{align}
The integration in $d^3q$ of Eq.~(\ref{ND}) is performed by using a cut-off $\Lambda\sim 700$ MeV for the modulus of the center of mass momentum of the $KZ_c(3900)$ system,  $\vec{q}^{\,*}$, in the triangular loops, i.e.,
\begin{align}
\int\frac{d^3q}{(2\pi)^3}\to\frac{1}{(2\pi)^2}\int\limits_0^\infty d|\vec{q}\,|\,|\vec{q}\,|^2\int\limits_{-1}^1 d\text{cos}\theta\,\Theta(|\vec{q}^{\,*}|-\Lambda).\label{Lamb}
\end{align}
Such value for $\Lambda$ corresponds to the one used when generating the $K^*(4307)$ from the $KD\bar D^*$ system, with the $D\bar D^*$ in isospin 1 system forming the $Z_c(3900)$~\cite{Ren:2018pcd}.  The vectors $\vec{q}$ and $\vec{q}^{\,*}$ in Eq.~(\ref{Lamb}) are related through a boost~\cite{Bayar:2016ftu}
\begin{align}
\vec{q}^{\,*}=\left[\left(\frac{E_{KZ(\mathcal{A})}}{m_{KZ(\mathcal{A})}}-1\right)\frac{\vec{q}\cdot\vec{p}_\mathcal{A}}{\vec{p}^{\,2}_\mathcal{A}}+\frac{\omega_K}{m_{KZ(\mathcal{A})}}\right]\vec{p}_\mathcal{A}+\vec{q},
\end{align}
with $m_{KZ(\mathcal{A})}$ being the invariant mass of the $KZ_c(3900)$ system in the triangular loop and $E_{KZ(\mathcal{A})}=\sqrt{m^2_{KZ(\mathcal{A})}+\vec{p}^{\,2}_\mathcal{A}}$ its energy in the rest frame of the decaying particle.
\bibliographystyle{unsrt}
\bibliography{refsB4307decay}

\end{document}